\documentclass[journal]{IEEEtran}

\ifCLASSINFOpdf
 
\else
  
\fi
\usepackage{cite}
\usepackage{amsthm}
\usepackage{graphicx}
\usepackage{epstopdf}
\usepackage{amsmath}
\usepackage{amsfonts}
\usepackage{amssymb}
\usepackage{xcolor}
\usepackage{multirow}
\usepackage{multicol}
\usepackage[utf8]{inputenc}
\usepackage[english]{babel}
\definecolor{orange}{RGB}{0,112,192}

\DeclareMathOperator*{\argmin}{arg\,min}
\DeclareMathOperator*{\argmax}{arg\,max}

\newtheorem{prop}{Proposition}

\usepackage[color=black,opacity=1]{background}

\SetBgContents{ \begin{minipage}{35cm}{\copyright 2023 IEEE. Personal use
      of this material is permitted. Permission from IEEE must be obtained
       for all other uses, in any current or future media, including
        reprinting/republishing this material for advertising or promotional purposes,
         creating new collective works, for resale or redistribution to servers or lists, 
         or reuse of any copyrighted component of this work in other works. This paper will appear in the IEEE   Transactions on Wireless Communications, 2023.}\end{minipage}}
\SetBgScale{.5}
\SetBgAngle{0}
\SetBgPosition{current page.south west}
\SetBgHshift{20cm}
\SetBgVshift{2cm}

\begin{document}
\makeatletter

\newcommand*{\rom}[1]{\expandafter\@slowromancap\romannumeral #1@}
\makeatother
\title{\huge Massive MIMO with Cauchy Noise: Channel Estimation, Achievable Rate and Data Decoding}

\author{Ziya~G\"{u}lg\"{u}n,
         and~Erik~G.~Larsson,~\IEEEmembership{Fellow,~IEEE} 
\thanks{Z. G\"{u}lg\"{u}n was with the Department
of Electrical Engineering (ISY), 58183 Link\"{o}ping, Sweden. He is now with Ericsson AB, 16440 Stockholm, Sweden (ziya.gulgun@ericsson.com).

E. G. Larsson is  with the Department
of Electrical Engineering (ISY), 58183 Link\"{o}ping, Sweden e-mail: (erik.g.larsson@liu.se).

This work was supported by Security Link and the SURPRISE project funded by the Swedish Foundation for Strategic Research (SSF). A preliminary version of this paper was presented at the International Conference on Communications (ICC), 2022 \cite{ICC2022}.
}}

\maketitle
 
\begin{abstract}
We consider  massive multiple-input multiple-output (MIMO) systems in the presence of  Cauchy noise. 
First, we focus on the channel estimation problem. In the standard massive MIMO setup, the users transmit orthonormal pilots during the training phase and the received signal at the base station is projected onto each pilot. This processing is optimum when the noise is Gaussian.
We show that this processing is not optimal when the noise is Cauchy and as a remedy  propose a channel estimation technique that operates on  the raw received signal. Second, we derive  uplink-downlink achievable rates in the presence of Cauchy noise for perfect and imperfect channel state information.
Finally, we derive  log-likelihood ratio expressions for soft bit detection for both  uplink and downlink, and simulate coded  bit-error-rate curves. In addition to this, we derive and compare the symbol detectors in the presence of both Gaussian and Cauchy noises. An important observation is that  the detector constructed for Cauchy noise performs well with both Gaussian and Cauchy noises; on the other hand, the detector for Gaussian noise works poorly in the presence of  Cauchy noise. That is, the Cauchy detector is robust against heavy-tailed noise, whereas the Gaussian detector is not.
\end{abstract}


\IEEEpeerreviewmaketitle

\section{Introduction}

Massive multiple-input multiple-output (MIMO) is one of the core technologies in the 5G physical layer \cite{BJORNSON20193}. 
A massive MIMO base station (BS) is equipped with a large number of antennas (on the order of 100), each connected to an independent radio-frequency chain. 
Thanks to  this flexibility, the BS can serve tens of users on the same time-frequency resources simultaneously.
During the last decade,  many aspects of massive MIMO technology  have been  investigated in depth, for example: 
 power allocation and user association algorithms   \cite{7497508}, hardware impairments \cite{8429913}.

This paper discusses an important aspect that has been largely ignored in the massive MIMO literature: the fact that noise and interference may not be Gaussian in general. More specifically, the noise can be impulsive and have a heavy-tailed distribution, corresponding to the presence of outliers. For example, in \cite{9356519}, impulsive noise was mentioned  as one of the physical-layer challenges for  6G.
\subsection{Motivation for Impulsive Noise}
The Gaussian noise assumption is justified if the noise results from  superposition of many independent components. However, this assumption does not apply in  many real situations that may occur in  electronic devices \cite{7763811}. For example, in \cite{meas} the authors performed a series of measurements of  electromagnetic noise for industrial wireless communications, and identified impulsive noise which can be modelled via Cauchy or Gamma distributions. Multi-carrier transmission with impulsive noise was studied in \cite{ofdm,8447433,6930962}. Adaptive demodulation for channels with  impulsive noise  was studied in \cite{9645175}. Other examples are that ambient noise in   shallow water for acoustic communication is highly impulsive \cite{1707997}; 
the noise in powerline communication channels is impulsive \cite{990732}; 
interference in ad hoc networks can be impulsive \cite{165447};
and  clutter models in radar signal processing are typically non-Gaussian \cite{7805183}. 
Moreover,  impulsive noise can also be generated by malicious transmitters (jammers).

One way to  model impulsive noise is as symmetric $\alpha$-stable  (S$\alpha$S) random variables. 
The smaller $\alpha$ is, the more impulsive the noise is. 
When $\alpha$ is $1$ and $2$, the S$\alpha$S distribution becomes  Cauchy and Gaussian, respectively. 
In the literature,  impulsive noise (S$\alpha$S) has been studied in  various contexts. 
For example,    array signal processing with   impulsive noise is treated  in \cite{510611,550156}. Spectrum sensing with S$\alpha$S channel was studied in \cite{8781820}.
Localization problems in the presence of $\alpha$-stable noise were investigated in \cite{482119}.
The probability density functions (pdfs) of  S$\alpha$S distributions are approximated and the signal detection performances of these pdfs  are investigated in \cite{6932454}.
In \cite{4374156}, soft-decision metrics for  coded signals in  Cauchy noise were derived. 
An interesting conclusion in \cite{4374156} is that the performance loss of the detector designed for  Cauchy noise when exposed to Gaussian noise is much smaller than the loss when the  detector designed for  Gaussian noise is exposed to Cauchy noise. 
Indeed,  the detector designed for  Gaussian noise works poorly in the presence of Cauchy noise, whereas the detector designed for Cauchy noise is very robust to the actual distribution of the noise. 

We consider massive MIMO  specifically   with Cauchy  (S$(\alpha=1)$S) noise. 
Another paper that addressed      channel estimation  for massive MIMO with impulsive noise is \cite{9291441}.  The impulsive noise in \cite{9291441} is modelled as a mixture of   Gaussian noise and   outliers, rather than via an S$\alpha$S distribution; the mixture weights (probabilities of outliers) are estimated by  tuning a   sparsity level parameter.  In contrast to \cite{9291441}, we chose to work with Cauchy noise since it is simpler to deal with analytically (for example, the Cauchy distribution is preserved under linear combinations) and it has fewer parameters that require tuning to the impulsive level of the noise. In addition, compared with \cite{9291441}, we provide achievable rate expressions and soft decoding metrics.

\subsection{Summary of Technical Contributions and Organization of the Paper}
We describe  a system model for single-cell massive MIMO with   Cauchy noise and focus on the channel estimation in Section~\ref{sec:sys and ch} where we derive two types of channel estimators. In the first, the received signal at the base station (BS) is de-spread by correlating it with each user's pilot signal. In the presence of Gaussian noise, this
entails no loss of information; the de-spread received pilots  are  sufficient statistics. 
However, this is not the case   if the noise is non-Gaussian.  
The second channel estimation approach  rather estimates the channels from the raw received signal without de-spreading first, and is shown to be superior. 
Next, in Section~\ref{sec:Ach_rate}, we derive  uplink and downlink achievable rate expressions for the cases of perfect and imperfect channel state information (CSI). Thereby, for the imperfect CSI case, the channel estimates are used as side information. 
In Section~\ref{sec:Data_dec}, we then   derive log-likelihood ratio (LLR) expressions for soft decoding both for uplink and downlink.
Finally, in Section~\ref{sec:Sim_res} we give numerical results on achievable rates and bit-error-rate (BER) for the different detectors.

This paper is a comprehensive extension of our conference paper  \cite{ICC2022}. 
The new material compared to   \cite{ICC2022} is mainly that (i)  we analyze  achievable rates for both perfect and imperfect CSI, (ii) we analyze the decoding performance of a Cauchy receiver in the presence of noise with other S$\alpha$S distributions, (iii) we derive LLR expressions for soft decoding, and (iv) we present numerical results on achievable rates and coded BER.

\subsection{Notation}

$|.|, (.)^T, (.)^*$ and $(.)^H$ denote the absolute value of a scalar, determinant of a matrix, transpose, conjugate and conjugate transpose operators, respectively.  Boldface lowercase letters, $\mathbf{x}$, denote column vectors, boldface uppercase letters, $\mathbf{X}$, denote matrices, and  uppercase letters, $X$, denote random variables. The $l_p$ norm is denoted by  $\|\mathbf{x}\|_p$. $\mathbb{E}[.]$ refers to the expectation operator. $\Re(.)$ is the real part of a complex number.

\section{System Model and Channel Estimation}
\label{sec:sys and ch}
\subsection{System Model}
\label{sec:System model}
We consider a single-cell massive MIMO system including a BS equipped with $M$ antennas, serving $K$ users where each user has a single antenna. 
A block-fading model is considered in which the channel is constant and frequency-flat in a coherence block with size $T$ samples. 
A sub-block with length $\tau$ is reserved for the channel estimation and the rest of the coherence block with length  $T-\tau$ is dedicated to  data transmission. 

During the training phase, the users transmit orthonormal pilot vectors with length $\tau$ to the BS. 
$\tau$ should be greater than or equal to $K$ and less than $T$. 
Let us denote the pilot vector for the $k^{th}$ user by $\boldsymbol{\phi}_k\in\mathbb{C}^{\tau}$. 
Since these vectors are orthonormal, $\|\boldsymbol{\phi}_k\|^2=1$ and $\boldsymbol{\phi}_k^H\boldsymbol{\phi}_j=0$ for $k\neq{j}$. 
It remains an open question under what exact circumstances orthogonal pilot sequences are optimal in the presence of Cauchy noise. (For example, in Cauchy noise, unlike in Gaussian noise, de-spreading is suboptimal; see Section~\ref{sec:channel_est}.) The choice of orthogonal pilots, however, is optimal for Gaussian noise and it is the standard design choice most systems; hence, we
adopt it here. 
Even when restricting the pilots coming from an orthogonal pilot book, 
numerical evidence shows that the performance is different for different choices of this pilot book (unlike in the Gaussian case). For example,   pilots chosen from the identity matrix and  pilots chosen from a normalized discrete Fourier transform (DFT) matrix do not provide the same performances. In this work, we choose the pilots from the normalized DFT matrix because this increases the number of observed realizations in the receiver side, which is important  because of the outliers in the Cauchy distribution.

The received pilot signal in the BS can be expressed as:
\begin{equation}
\label{eq:rec}
\mathbf{Y}=\sum_{k=1}^K\sqrt{{\tau}p_k}\mathbf{h}_k\boldsymbol{\phi}_k^T+\mathbf{N},
\end{equation}
where $p_k$ is the power of the received signal corresponding to the $k^{th}$ user, and $\mathbf{h}_k\in\mathbb{C}^M$ is the $k^{th}$ user's channel vector. 
The elements in all channel vectors are independent, identically distributed (i.i.d.) circularly symmetric complex Gaussian random variables, i.e., $\mathbf{h}_k\sim\mathcal{CN}(0,\mathbf{I})$, and $\mathbb{E}\{\mathbf{h}_k^H\mathbf{h}_m\}=0$ for $k\neq{m}$, corresponding to uncorrelated Rayleigh fading. 
$\mathbf{N}$ is an $M\times{\tau}$ noise matrix containing i.i.d.  isotropic  complex Cauchy random variables with   dispersion parameter $\gamma=1$. 
If the noise components were  complex Gaussian  with unit variance, $p_k$ would have the meaning of  signal-to-noise ratio (SNR). 
However, it is not meaningful to define $p_k$ as SNR with Cauchy noise, because the second-order moment of the Cauchy distribution is infinite; see Appendix~\ref{sec:Math_pr}. 
Therefore, we use the term signal-to-dispersion ratio (SDR)  for $p_k$   in this work. 

To contrast SDR and SNR, we investigate the relation between the standard deviation of the Gaussian distribution and the dispersion of the Cauchy distribution. For the (real-valued) Gaussian distribution, the probability that a realization, $X$, lies between $-\sigma$ and $\sigma$, where $\sigma$ is the standard deviation, is around 2/3.
The probability that a  (real-valued) Cauchy realization, $X$, lies between $0$ (the median) and $t$ can be expressed as:
\begin{equation}
\label{eq:Cau}
P(0<X<t)=\frac{1}{\pi}\tan^{-1}\left(\frac{t}{\gamma}\right),
\end{equation}
where $\gamma$ is the dispersion.
From \eqref{eq:Cau}, we infer that the probability that a Cauchy realization, $X$, lies between
$-1.7\gamma$ and $1.7\gamma$ is around 2/3.
Therefore, if we set $\sigma=1.7\gamma$, the probability is 2/3 for both a Gaussian and a
Cauchy random variable. 

 \subsection{Channel Estimation} 
\label{sec:channel_est}
In this section, we propose two channel estimation techniques for the case when  Cauchy noise is present. The first relies on de-spreading of the   received pilots. The second, in contrast, operates on  the original received pilots in \eqref{eq:rec} without performing de-spreading.

\subsubsection{Channel Estimation with De-spreading Operation}

After the de-spreading operation, the signal corresponding to the $k^{th}$ user can be expressed as:
\begin{equation}
\label{eq:despread}
\mathbf{y}_k=\mathbf{Y}\boldsymbol{\phi}_k^*=\sqrt{{\tau}{p}_k}\mathbf{h}_k+\mathbf{n}_k,
\end{equation}
where $\mathbf{n}_k=\mathbf{N}\boldsymbol{\phi}_k^*$. Now, let us analyze the statistic of $\mathbf{n}_k$.
\begin{prop}
$\mathbf{n}_k$ contains i.i.d. isotropic complex Cauchy random variables with the dispersion $\gamma_k^{\prime}=\gamma\|\boldsymbol{\phi}_k\|_1=\gamma\displaystyle\sum_{i=1}^{\tau}|{\phi}_k[i]|$ where ${\phi}_k[i]$ is the $i^{th}$ element of $\boldsymbol{\phi}_k$.
\end{prop}
\begin{IEEEproof}
First, since the row vectors in $\mathbf{N}$ are independent, the elements in $\mathbf{n}_k$ are mutually independent. 
 
Next, as a preliminary observation that we will exploit in the proof, we analyze the statistics of $cX$ where $c\in\mathbb{C}$ and $X$ is an isotropic complex Cauchy random variable. 
By using \eqref{eq:char}, we can express the characteristic function of $cX$ as:
\begin{equation}
\label{eq:firs}
\phi_{(cX)}(\omega)=\mathbb{E}\left[\exp(j\Re(\omega{(cX)^*}))\right]=\mathbb{E}\left[\exp(j\Re(c^*\omega{X^*}))\right].
\end{equation}
Let us define $\omega^{\prime}=c^*\omega$. Then  \eqref{eq:firs} can be rewritten as:
\begin{equation}
\label{eq:fund}
\mathbb{E}\left[\exp(j\Re(\omega{'}{X^*}))\right]=\exp(-\gamma|\omega{'}|)=\exp(-\gamma|c||\omega|).
\end{equation}
Now, we focus on the $l^{th}$ element in $\mathbf{n}_k$, denoted as $\mathbf{n}_k[l]$. The characteristic function of $\mathbf{n}_k[l]$ is:
\begin{subequations}
\begin{align}
\phi_{\mathbf{n}_k[l]}(\omega)&=\mathbb{E}\left[\exp(j\Re(\omega{(\mathbf{n}_k[l])^*}))\right] \\ 
\label{eq:ind}
&=\mathbb{E}\left[\exp\left(j\Re\left(\omega{\sum_{i=1}^{\tau}\boldsymbol{\phi}_k[i]\mathbf{N}[l,i]^*}\right)\right)\right] \\
\label{eq:ind_2} 
&=\prod_{i=1}^\tau\mathbb{E}\left[\exp(j\Re(\omega{\boldsymbol{\phi}_k[i]\mathbf{N}[l,i]^*}))\right],
\end{align}
\end{subequations}
where  \eqref{eq:ind} can be represented as a product of random variables as in \eqref{eq:ind_2} because all random variables are mutually independent ($\mathbf{N}[l,i]$ is the element in the $l^{th}$ row and the $i^{th}$ column of $\mathbf{N}$ and $\mathbf{N}$ contains i.i.d. random variables). By using \eqref{eq:fund}, the following is obtained:
\begin{align}
\phi_{\mathbf{n}_k[l]}(\omega)&=\prod_{i=1}^\tau\mathbb{E}\left[\exp(j\Re(\omega{\boldsymbol{\phi}_k[i]\mathbf{N}[l,i]^*}))\right] \\ \nonumber
&=\prod_{i=1}^{\tau}\exp(-\gamma|\boldsymbol{\phi}_k[i]||\omega|) \\ \nonumber 
&=\exp\left(-\gamma\sum_{i=1}^{\tau}|\boldsymbol{\phi}_k[i]||\omega|\right),
\end{align}
which concludes the proof.
\end{IEEEproof}

Proposition 1 presents an important result. 
Although the de-spreading operation preserves the distribution  of noise, the dispersion may increase. 
For example, if the pilot signals are chosen from a normalized DFT matrix, the dispersion after de-spreading operation becomes $\gamma_k^{\prime}=\gamma\sqrt{\tau} \ \forall k$. 
This is not the case for the isotropic complex Gaussian noise  because a   multiplication with a unitary matrix does not change the noise variance \cite[Chapter~3]{marzetta_larsson_yang_ngo_2016}.

Now, we need to estimate the channels from \eqref{eq:despread}. 
Most existing papers use minimum mean-square-error (MMSE) or linear MMSE techniques to estimate the channels \cite{marzetta_larsson_yang_ngo_2016,massivemimobook}. 
However, these techniques cannot be applied in the presence of  Cauchy noise because the first and second order moments for this noise distribution are undefined. 
Therefore, here we take a   maximum-likelihood (ML) approach. 
The maximum-a-posteriori (MAP) technique could also be applied if a prior on the channel statistic is known. 
The  difference between MAP and ML  is   that the pdf of channel statistic appears in the objective function;
in this paper, we only consider  ML for simplicity. 

From   \eqref{eq:despread}, the likelihood function for $\mathbf{h}_k$ based on the de-spread data is 
\begin{equation}
\label{eq:ML_cost}
p(\mathbf{y}_k|\mathbf{h}_k)=\prod_{i=1}^M\frac{\gamma_k'}{2\pi((\gamma_k')^2+|\mathbf{y}_k[i]-\sqrt{{\tau}{p}_k}\mathbf{h}_k[i]|^2)^{3/2}}.
\end{equation}
Based on   \eqref{eq:ML_cost}, the ML   estimate of the channel $\mathbf{h}_k$ is:
\begin{equation}
\label{eq:ML}
\hat{\mathbf{h}}_k^\text{ML}=\argmax_{\mathbf{h}_{k}}p(\mathbf{y}_k|\mathbf{h}_k)=\frac{\mathbf{y}_k}{\sqrt{{\tau}{p}_k}}.
\end{equation}
From  \eqref{eq:despread} and \eqref{eq:ML}, it can be immediately seen that $\hat{\mathbf{h}}_k^\text{ML}$ is also identical with the  least-squares channel estimate for $\mathbf{h}_k$.

Next, we analyze  whether $\mathbf{y}_k$ is a sufficient statistic or not. To do this,   we go back the unprocessed received signal in \eqref{eq:rec}. 
The likelihood function based on the raw data in \eqref{eq:rec} is:
\begin{align} 
\label{eq:general_ML}
&p\left(\mathbf{Y}|\mathbf{h}_1,\hdots,\mathbf{h}_K\right)= \\ \nonumber
&\prod_{l=1}^M\prod_{i=1}^\tau\frac{\gamma}{2\pi\left(\gamma^2+|\mathbf{Y}[l,i]-\sum_{k=1}^K\sqrt{{\tau}{p}_k}\mathbf{h}_k[l]\boldsymbol{\phi}_k[i]|^2\right)^{3/2}},
\end{align}
where $\mathbf{Y}[l,i]$ is the element in the $l^{th}$ row and $i^{th}$ column of $\mathbf{Y}$. In order to obtain a sufficient statistic, there should exist a factorization of the likelihood function in \eqref{eq:general_ML}  as \cite[Chapter~4]{Kay:1993:FSS:151045}:
\begin{equation}
p\left(\mathbf{Y}|\mathbf{h}_1,\hdots,\mathbf{h}_K\right)=h(\mathbf{Y})g(T(\mathbf{Y}),\mathbf{h}_1,\hdots,\mathbf{h}_K),
\end{equation}
where $g(\cdot)$ and $h(\cdot)$ are some  functions, and $T(\cdot)$ is the sufficient statistic. Note that $g(\cdot)$ should depend on $\mathbf{Y}$ only through $T(\mathbf{Y})$. Because of the fractional exponent $3/2$ in the denominator, one cannot expand the corresponding term. 
One can multiply the all the terms under the fractional exponent 3/2 but a function $T(\cdot)$  cannot be obtained because of the many cross-terms that appear.

\emph{Remark~1}:
Let us consider an affine estimator of the $k^{th}$ user's channel which has the structure  $\mathbf{Y}\mathbf{a}_k+\mathbf{b}_k$ where $\mathbf{Y}\mathbf{a}_k\neq{0}$, $\mathbf{a}_k\in\mathbb{C}^{\tau}$ and  $\mathbf{b}_k\in\mathbb{C}^{M}$. As a result of Proposition 1, this affine estimator can be linearly decomposed into noise which has a complex isotropic Cauchy distribution and a signal part. For example, from \eqref{eq:despread},   $\mathbf{n}_k$ is the additive part of $\mathbf{y}_k$ which has the complex isotropic Cauchy distribution. Therefore, the mean of $\mathbf{y}_k$ is undefined and the variance of $\mathbf{y}_k$ is infinite for the affine estimators.

\subsubsection{Channel Estimation from the Unprocessed Received Signal}
\label{sec:est_unprocess}
In this section, we derive the ML estimates for the channel vectors based on the unprocessed received signal in  \eqref{eq:rec}. 
The likelihood function is given in \eqref{eq:general_ML}. 
Maximizing \eqref{eq:general_ML} is very complicated   because the parameters of all users interact with each other. 
To solve this problem, one approach is to use a coordinate search algorithm \cite[Chapter~9]{nocedal2006numerical}. 
The idea is as follows: First, assign some initial values to all parameters except  $\mathbf{h}_1$. Then find the estimate of $\mathbf{h}_1$ that maximizes the likelihood function. 
Next,  with the so-obtained value of  $\mathbf{h}_1$,   find the estimate of $\mathbf{h}_2$ that maximizes the likelihood function and so forth. 
One can then iteratively loop through all   $\mathbf{h}_k$ and then return to $\mathbf{h}_1$, and run another complete iteration over all $k$. 
This procedure goes on until there is a sufficiently small  difference between the norms of the   channel estimates in the current round compared to in the   previous round. 

The coordinate descent algorithm guarantees that the function value obtained in an iteration is less than or equal to the function value obtained in the previous iteration. However, the algorithm may converge to a non-stationary point, or the algorithm  can continue searching infinitely many times for non-convex functions \cite{Powell1973OnSD}. The examples in \cite{Powell1973OnSD} are very special examples, but the conclusions therein prevent us from  making a general statement that the algorithm converges to a stationary point.  Bertsekas \cite{bertsekas2016nonlinear} shows in Proposition 2.7.4 that if the objective is continuously differentiable, i.e., the first derivative is continuous, and the minimizer along any coordinate direction from any point is unique, then the algorithm converges to a stationary point. The objective function in \eqref{eq:general_ML} is continuously differentiable, but we do not have proof that the minimizer along any direction is not unique. (In \cite{ICC2022}, we  claimed that the coordinate descent algorithm converges to a locally optimum point, but this is unknown.)
Numerical experiments suggest that the convergence to a stationary point depends on the initial point.

In the first iteration, we assign $\hat{\mathbf{h}}_k^\text{ML}$ in \eqref{eq:ML} to $\mathbf{h}_k$ for $k=2,\hdots,K$ as initial solutions. 
This results in the modified  objective function in  \eqref{eq:general_ML} for $\mathbf{h}_1$ as follows:
{\small{\begin{equation}
\label{eq:ML_h}
p(\mathbf{Y'}|\mathbf{h}_1)=\prod_{l=1}^M\prod_{i=1}^\tau\frac{\gamma}{2\pi\left(\gamma^2+|\mathbf{Y'}[l,i]-\sqrt{{\tau}{p}_1}\mathbf{h}_1[l]\boldsymbol{\phi}_1[i]|^2\right)^{3/2}},
\end{equation}}}
where
\begin{equation}
\nonumber
 \mathbf{Y'}=\mathbf{Y}-\sum_{k={2}}^K\sqrt{{\tau}{p}_{k}}\hat{\mathbf{h}}_k^\text{ML}\boldsymbol{\phi}_{k}^{T}.
\end{equation}
When maximizing with respect to  the first element of $\mathbf{h}_1$, we have $M$ separable maximization problems. 
The corresponding observation vector is  the first row from $\mathbf{Y'}$. Therefore, we need to maximize the following:
\begin{equation}
\label{eq:mod}
\prod_{i=1}^\tau\frac{\gamma}{2\pi\left(\gamma^2+|\mathbf{Y'}[1,i]-\sqrt{\tau{p}_1}\mathbf{h}_1[1]\boldsymbol{\phi}_1[i]|^2\right)^{3/2}}.
\end{equation}
Taking the logarithm, the maximization of \eqref{eq:mod} is equivalent to the minimization of the following:
\begin{equation}
\label{obj}
\sum_{i=1}^\tau\log\left(\gamma^2+|\mathbf{Y'}[1,i]-\sqrt{\tau{p}_1}\mathbf{h}_1[1]\boldsymbol{\phi}_1[i]|^2\right).
\end{equation}
To minimize \eqref{obj}, we use the gradient descent algorithm \cite[Chapter~3]{nocedal2006numerical} and give the details in Appendix~\ref{sec:opt_pro}. Note that the function in \eqref{obj} is nonconvex because the Cauchy distribution is non-log-convex.

\emph{Remark~2:}
Suppose we want to estimate a parameter from data containing Cauchy noise. The estimate may have a certain variance and even the Cramer-Rao bound may exist for the estimate \cite{482119}. In general, when the noise is Cauchy, the relation between the ML estimate  and the observed data is not affine, not even for a linear signal model. Hence, the ML estimate may have a certain variance, although sometimes it does not; for example, see \eqref{eq:ML}.

\section{Achievable Rates with Cauchy Noise}
\label{sec:Ach_rate}
In this section, we present the uplink and downlink achievable rates for the massive MIMO communication link with perfect and imperfect CSI. In the literature,  \cite{6875400} derived the capacity for the scalar Cauchy channel under a logarithmic constraint on the input distribution; see \cite{6875400} for details. Another paper \cite{7866884} derived some capacity bounds for S$\alpha$S noise channel with $\alpha>1$ which does not cover the Cauchy noise. 
As the exact capacity appears intractable for our setup when the noise is Cauchy, in this paper we focus on achievable rates (lower bounds on capacity).
To study the effects of the Cauchy noise specifically, we restrict this analysis to the  one-user scenario.

\subsection{The Uplink Achievable Rate}
\label{up_ach}
The mutual information of the single-input single-output (SISO) channel can be calculated empirically;   see   details in Appendix \ref{sec:mut_inf}. Now we apply  this result to the case of  a single-input multiple-output channel with Cauchy noise.  
We start with the perfect CSI case.
Mathematically, we can express the received signal as an $M$-vector:
\begin{equation}
\mathbf{y}=\sqrt{p^{ul}}\mathbf{h}x+\mathbf{n},
\end{equation}
where $p^{ul}$ is the uplink SDR, $x$ is the transmitted signal, $\mathbf{h}\in\mathbb{C}^M$ is the channel gain. Both the channel gain and its statistics are known by the receiver. $\mathbf{n}\in\mathbb{C}^M$ is the noise vector including complex isotropic Cauchy random variables with unit dispersion parameter. Let us define a joint random variable $\bar{Y}=[Y_1,\hdots,Y_M]^T$ comprising the random variables observed at  each antenna and a joint random variable $\bar{H}=[H_1,\hdots,H_M]^T$ including the channel realizations known by the receiver. Assume that the transmitted signal is chosen uniformly from the discrete set $A_X$ with cardinality $S$. After this, we can define the uplink achievable rate:
\begin{equation}
\label{gene}
R^{ul}=\log_2{S}-\mathbb{E}_{X,\bar{Y},\bar{H}}\left\{\log_2\left(\frac{\displaystyle\sum_{x\in{A}_X}p(\bar{Y},\bar{H}|x)}{p(\bar{Y},\bar{H}|X)}\right)\right\}.
\end{equation}
All $(Y_i,H_i)$ pairs are independent to each other if $X$ is given. Hence, \eqref{gene} can be rewritten as:
\begin{align}
R^{ul}&=\log_2{S}-\mathbb{E}_{X,\bar{Y},\bar{H}}\left\{\log_2\left(\frac{\displaystyle\sum_{x\in{A}_X}\prod_{i=1}^Mp(Y_i,H_i|x)}{\displaystyle\prod_{i=1}^Mp(Y_i,H_i|X)}\right)\right\} \\ \nonumber
&=\log_2{S}-\mathbb{E}_{X,\bar{Y},\bar{H}}\left\{\log_2\left(\frac{\displaystyle\sum_{x\in{A}_X}\prod_{i=1}^Mp(Y_i|H_i,x)}{\displaystyle\prod_{i=1}^Mp(Y_i|H_i,X)}\right)\right\},
\end{align}
where $p(y_i|h_i,x)$ is:
\begin{equation}
\label{eq:up_den}
p(y_i|x)=\frac{\gamma}{2\pi(\gamma^2+|y_i-\sqrt{p^{ul}}h_ix|^2)^{3/2}},
\end{equation}
where $y_i$ and $h_i$ are the $i^{th}$ elements of $\mathbf{y}$ and $\mathbf{h}$, respectively.

Now consider the imperfect CSI case. The BS estimates the channel denoted by $\mathbf{\hat{h}}$ in the training phase. These estimates are used as side information to calculate achievable rates. This calculation exploits standard results on capacity
bounding with side information; for example, see \cite[Sec.~2.3]{marzetta_larsson_yang_ngo_2016}. Since there exists an error between the estimates and the real channels, the achievable rates for the imperfect CSI case are less than or equal to the achievable rates for the perfect CSI case \cite{841172}. Let us define a joint random variable $\bar{\hat{H}}=[\hat{H}_1,\hdots,\hat{H}_M]^T$ including the random variables that represent the channel estimates. Then, the uplink achievable rate for the imperfect CSI case can be expressed as:
{\small{\begin{align}
\label{eq:rate}
&R^{ul, \text{imCSI}}=\left(1-\frac{\tau}{T}\right)\times
\\ \nonumber
&\left(\log_2{S}-\mathbb{E}_{X,\bar{Y},\bar{\hat{H}}}\left\{\log_2\left(\frac{\displaystyle\sum_{x\in{A}_X}\prod_{i=1}^Mp(Y_i|x, H_i=\hat{H}_i)}{\displaystyle\prod_{i=1}^Mp(Y_i|X, H_i=\hat{H}_i)}\right)\right\}\right),
\end{align}}}
where $p(y_i|x, h_i=\hat{h}_i)$ can be obtained from \eqref{eq:up_den} when $h_i=\hat{h}_i$.
The pre-log factor  appears in \eqref{eq:rate} because we do not send any data during $\tau$ samples of each coherence block. 

It is important to note that we do not have the statistics of the channel estimation error, because we do not have  closed-form channel estimates. However, for the uplink the BS can calculate some statistics of the channel estimation error empirically. In \ref{sec:ch_est_err}, we explain how the BS calculates the statistics of the channel estimation error and  how the
effects of this estimation error can be considered as additional noise term.

\emph{Remark~3}: Let us define a variable $y=\mathbf{v}^H\mathbf{y}$, where $\mathbf{v}$ is a decoding vector. For the perfect CSI and the Gaussian cases, we do not lose any information if $\mathbf{v}=\mathbf{h}/\|\mathbf{h}\|$ \cite{Tse05fundamentalsof}. In other words, $y$ is a sufficient statistic for $\mathbf{y}$. However this is not the case in Cauchy noise. Therefore, we derive the rate expressions by using $\mathbf{y}$ for perfect and imperfect CSI.

\subsection{The Downlink Achievable Rate}
\label{sec:down}
Now we find the achievable rate of a communication link where the BS transmits the data to the user. Again, we start with the perfect CSI case.
The received signal can be written as, 
\begin{equation}
y=\sqrt{p^{dl}}\mathbf{h}^T\mathbf{a}x+n,
\end{equation}
 where $p^{dl}$ is the downlink SDR, $x$ is the transmitted signal, $\mathbf{h}\in\mathbb{C}^M$ is the channel gain that is known by the receiver, $\mathbf{a}\in\mathbb{C}^M$ is the precoder vector designed by the BS, $n$ is the Cauchy noise with unit dispersion parameter. We assume that $\|\mathbf{a}\|_2=1$ so that we do not obtain any power gain from the precoder. 
 To maximize the received useful signal power,  $\mathbf{a}$ should equal $\mathbf{h}^{*}/\|\mathbf{h}\|$. 
 Hence, the downlink achievable rate becomes:
\begin{equation}
\label{dl}
R^{dl}=\log_2{S}-\mathbb{E}_{X,Y}\left\{\log_2\left(\frac{\displaystyle\sum_{x\in{A}_X}p({Y}|x)}{p({Y}|X)}\right)\right\},
\end{equation}
where $p(y|x)$ is:
\begin{equation}
\label{eq:down_den}
p(y|x)=\frac{\gamma}{2\pi(\gamma^2+|y-\sqrt{p^{dl}}\|\mathbf{h}\|_2x|^2)^{3/2}}.
\end{equation}
Similar to the uplink achievable rates, we use the channel estimates as side information to calculate the downlink achievable rates. To do this,  the precoder vector should be $\mathbf{\hat{h}}^{*}$ for the imperfect CSI case. Let us define a joint random variable $\bar{H}=[H_1,\hdots,H_M]^T$. The achievable rate is given by:
\begin{align}
\label{eq:dl_rate}
&R^{dl, \text{imCSI}}=\left(1-\frac{\tau}{T}\right) \times
\\ \nonumber
&\left(\log_2{S}-\mathbb{E}_{X,{Y},\hat{\bar{H}}}\left\{\log_2\left(\frac{\displaystyle\sum_{x\in{A}_X}p(Y|x, \bar{H}=\hat{\bar{H}})}{p(Y|X,  \bar{H}=\hat{\bar{H}})}\right)\right\}\right),
\end{align}
where $p(y|x,  \mathbf{h}=\mathbf{\hat{{h}}})$ can be obtained from \eqref{eq:down_den} when $\mathbf{h}=\mathbf{\hat{{h}}}$.

In the downlink, the effect of the  channel estimation error is implicit,   entering only via  the precoder vector, but not explicitly
visible in the achievable rate expression.

\emph{Remark~4:} For the Gaussian noise and perfect CSI case, we observe the uplink-downlink duality when the maximum-ratio (MR) decoder and MR precoder are used in the uplink and downlink, respectively. Therefore, we have the same achievable rates. However, this appears not to be the case with Cauchy noise, because linear processing of the received signal in the uplink is suboptimal. 

\subsection{The Cauchy Decoder with Other S$\alpha$S Noise}
In this section, we evaluate the performance of the Cauchy decoder in the presence of other complex isotropic S$\alpha$S noise. We focus on  complex isotropic S$\alpha$S noise with $1<\alpha{<}2$, for which   capacity bounds exist in  closed form \cite{7866884}. 

For  the complex isotropic S$\alpha$S noise with $1<\alpha{<}2$, the pdf can be expressed as  \cite[Chapter~3]{Panagi}:
\begin{equation}
\label{eq:com}
f_X(x)=\frac{1}{\pi\alpha\gamma^{2/\alpha}}\sum_{k=0}^{\infty}\frac{(-1)^{k}}{2^{2k+1}(k!)^2}\Gamma\left(\frac{2k+2}{\alpha}\right)\left(\frac{|x|}{\gamma^{1/\alpha}}\right)^{2k},
\end{equation}
where $\Gamma(\cdot)$ is the standard Gamma function. The pdf in \eqref{eq:com} is not in closed-form. Therefore, to develop a decoding metric, either this pdf would have to be  approximated, or
implemented via a table lookup. We consider that  the Cauchy decoding metric is used, and next evaluate how it performs when it is exposed to noise with  a  S$(1<\alpha{<}2)$S distribution.

Consider a SISO channel:
\begin{equation}
\label{eq:mis_rec}
y=\sqrt{p}x+n,
\end{equation}
where $y$, $x$ and $p$ are the received signal, transmitted signal and the received power, respectively. Let us assume that $\mathbb{E}[|X^R|]=\mathbb{E}[|X^I|]$. The capacity bounds presented in Eq. (3) 
of  \cite{7866884} apply for the case of  real-valued S$\alpha$S distributions. 
For complex  S$(1<\alpha{\leq}2)$S, the real and imaginary parts are statistically dependent, which means that the capacity \emph{at least} doubles when considering the real and imaginary parts of the
channel together. This results in the    bound:
\begin{equation}
\label{eq:Cap_bound}
C{\geq}\frac{2}{\alpha}\log_2\left(1+\left(\frac{\sqrt{p}c}{\mathbb{E}[|N^R|]}\right)^{\alpha}\right), \quad  \mathbb{E}\left[\big|X^R\big|\right]{\leq}c,
\end{equation}
where $C$ is the capacity for \eqref{eq:mis_rec}, and $N^R$  is the real part of noise having  the S$\alpha$S distribution for the real-valued realizations. Note that $N^R$ is identically distributed with the imaginary part. The fractional moment of $N^R$  can be expressed as \cite[Chapter~3]{Panagi}:
\begin{equation}
\mathbb{E}\left[\big|N^R\big|^p\right]=\gamma^{p/\alpha}\frac{2^{p+1}\Gamma\left(\frac{p+1}{2}\right)\Gamma\left(-\frac{p}{\alpha}\right)}{\alpha\sqrt{\pi}\Gamma\left(-\frac{p}{2}\right)}, \quad 0<p<\alpha.
\end{equation}

Now,  assume that $y$ in \eqref{eq:mis_rec} is decoded by using the Cauchy model. This decoding technique is called \emph{mismatched decoding} because the actual present noise has different statistics than what the decoder assumes  \cite{340469}. The achievable rate obtained from the mismatched decoder is called as  \emph{mismatched achievable rate}. The mismatched achievable rate is a lower bound on the actual mutual information \cite[eq. (34)]{1661831} so for the capacity:
\begin{equation}
\label{eq:mis_rate}
C{\geq}I(X;Y){\geq}\log_2S-\mathbb{E}_{X,Y}\left\{\log_2\frac{\sum_{x\in{A_X}}\tilde{p}(Y|x)}{\tilde{p}(Y|X)}\right\},
\end{equation}
where $\tilde{p}(y|x)$ is the decoder metric that is assumed to be Cauchy model.

In Section~\ref{sec:Sim_res}, we compare the capacity bound in \eqref{eq:Cap_bound} and the mismatched achievable rate in \eqref{eq:mis_rate}.

\section{Data Decoding}
\label{sec:Data_dec}
In this section, we present hard decision metrics for uncoded symbol detection and soft
decision metrics for coded bit detection, respectively. 
For the hard  symbol detection, we only consider the uplink for both Gaussian and Cauchy noises. 
For the soft bit  detection, we consider  the uplink and the downlink for  Cauchy case only.

\subsection{Symbol Detection for  The Uplink (Uncoded Modulation)}

In this section, we use the channel estimates obtained in Section~\ref{sec:channel_est} to infer the transmitted data. Express  the received signal as:
\begin{equation}\label{eq:symbolmodel}
\mathbf{r}=\sum_{k=1}^K\sqrt{p_k}\mathbf{h}_ks_k+\mathbf{n},
\end{equation}
where $s_k$ is the symbol transmitted  by  user $k$,  chosen from a certain alphabet. 
If the noise is Gaussian, the detector will be:
\begin{equation}
\label{eq:min_Gauss}
\mathbf{\hat{s}}=\argmin\limits_{\mathbf{s}}\sum_{i=1}^M\left|\mathbf{r}[i]-\sum_{k=1}^K\sqrt{{p}_k}\hat{\mathbf{h}}_k[i]s_k\right|^2.
\end{equation}

For the case of Cauchy noise in $\mathbf{n}$, we insert the  channel estimates into the likelihood function associated with (\ref{eq:symbolmodel}):
\begin{align}
\label{eq:tra_dens}
&p\left(\mathbf{r}|\hat{\mathbf{h}}_1,\hdots,\hat{\mathbf{h}}_K,\mathbf{s}\right)= \\ \nonumber
&\prod_{i=1}^M\frac{\gamma}{2\pi\left(\gamma^2+|\mathbf{r}[i]-\sum_{k=1}^K\sqrt{{p}_k}\hat{\mathbf{h}}_k[i]s_k|^2\right)^{3/2}},
\end{align}
where $\mathbf{s}=[s_1\hdots{s}_K]^T.$ Therefore the ML estimates of the symbols are:
\begin{equation}
\label{eq:min}
\mathbf{\hat{s}}=\argmin\limits_{\mathbf{s}}\sum_{i=1}^M\log\left(\gamma^2+\left|\mathbf{r}[i]-\sum_{k=1}^K\sqrt{{p}_k}\hat{\mathbf{h}}_k[i]s_k\right|^2\right).
\end{equation}

The problems in   \eqref{eq:min_Gauss} and \eqref{eq:min}  are  constellation-constrained minimization problems. 
Neglecting the constellation constraint, the solution of the problem in \eqref{eq:min_Gauss} is the well-known zero-forcing (ZF) detector:
\begin{equation}
\hat{\mathbf{s}}=(\mathbf{\hat{H}}^{{H}}\mathbf{\hat{H}})^{-1}\mathbf{\hat{H}}^H\mathbf{r},
\end{equation}
where $\mathbf{\hat{H}}\in\mathbb{C}^{M\times{K}}$ is a matrix where the $k^{th}$ column is $\mathbf{\hat{h}}_k$. 

For the problem with Cauchy noise, \eqref{eq:min}, again we neglect the constellation constraint and   obtain a soft decision for each symbol from the alphabet by evaluating the  minimum Euclidean distance. The problem in \eqref{eq:min}  can be solved by using the gradient descent algorithm that is described in Section \ref{sec:est_unprocess}. For the initial solution of $\mathbf{\hat{s}}$ in the gradient descent, 
we take the zero vector. 

Note that the problem in \eqref{eq:min} has more than one local optimum solution. The global solution of \eqref{eq:min_Gauss}, however, is   unique and easily given by the ZF.

\subsection{Soft Decision Metric for The Uplink (Coded Modulation)}
\label{bit_det}

Consider again the received signal in \eqref{eq:symbolmodel}. We now focus on the LLR expression for the $i^{th}$ bit of the $k^{th}$ symbol. This LLR   can be written as:
\begin{equation}
\label{llr:cau}
l_{k,i}=\log\left(\frac{p(b_{k,i}=0|\mathbf{r})}{p(b_{k,i}=1|\mathbf{r})}\right)=\log\left(\frac{\displaystyle\sum_{\mathbf{s}:b_{k,i}(\mathbf{s})=0}p(\mathbf{r}|\mathbf{s})p(\mathbf{s})}{\displaystyle\sum_{\mathbf{s}:b_{k,i}(\mathbf{s})=1}p(\mathbf{r}|\mathbf{s})p(\mathbf{s})}\right).
\end{equation}
The notation $\mathbf{s}:b_{k,i}(\mathbf{s})=\beta$ represents  that the set contains all $\mathbf{s}$ where
$b_{k,i}$ is equal to $\beta$. Without loss of  generality, we assume that $p(\mathbf{s})$ equals $1/S^K$, where $S$ is the cardinality of the symbol alphabet. The BS makes use of the channel estimates as the real channels. Therefore $p(\mathbf{r}|\mathbf{s})$ is the same as \eqref{eq:tra_dens}. 

The complexity of \eqref{llr:cau} is huge because of the many terms in the summation. To simplify the expression in  \eqref{llr:cau}, one approach is to replace the summation with the largest term; this  is called the \emph{max-log} approximation \cite{4520145}. However, it has still high complexity because we need to search through $\frac{S}{2}\times{S^{K-1}}$ candidates just for one bit. Therefore, finding the largest term may be infeasible. Instead, we propose the following:
\begin{itemize}
\item For the $k^{th}$ symbol, there are $S/2$ symbols whose $i^{th}$ bit is 0. Let us denote these symbols by $S_k^{i=0}=\{s_{k,1}^{i=0},\hdots,s_{k,S/2}^{i=0}\}$. For each $S/2$ symbol, we can solve the following maximization problem by neglecting the constellation constraint:
\begin{equation}
\label{eq:Prob}
\mathbf{\tilde{s}}_{k,t}^{i=0}=\argmax_{\mathbf{\tilde{s}}}p(\mathbf{r}|\mathbf{\tilde{s}},s_k^{i=0}=s_{k,t}^{i=0}),
\end{equation}
where $t$ ranges from $1$ to $S/2$, $\mathbf{\tilde{s}}_{k,t}^{i=0}$ is an $(K-1)\times{1}$ vector including soft estimates when $s_k^{i=0}=s_{k,t}^{i=0}$. By combining \eqref{eq:min} and \eqref{eq:Prob}, the problem in \eqref{eq:Prob} is equivalent to:
\begin{align}
\label{eq:min_eqv}
\mathbf{\tilde{s}}_{k,t}^{i=0}=\argmin\limits_{\mathbf{\tilde{s}}}\sum_{i=1}^M\log\bigg(&\gamma^2+\bigg|\mathbf{r}[i]-\sqrt{{p}_k}\hat{\mathbf{h}}_k[i]s_{k,t}^{i=0} \\ \nonumber
-&\sum_{n=1, n\neq{k}}^K\sqrt{{p}_n}\hat{\mathbf{h}}_n[i]\tilde{s}_n\bigg|^2\bigg).
\end{align}

We can obtain hard estimates from  $\mathbf{\tilde{s}}_{k,t}^{i=0}$ based on the Euclidean distance between soft estimates and the real symbols. Let us denote  $\mathbf{\hat{s}}_{k,t}^{i=0}$ a vector including the hard estimates.
\item Choose the maximum among the $S/2$ likelihood terms. The LLR is approximately equal to:
\begin{align}
l_{k,i}\approx&\log\left({\max_{\{{\mathbf{\hat{s}}}_{k,t}^{i=0}\}_{t=1}^{S/2}}p(\mathbf{r}|\mathbf{\hat{s}}_{k,t}^{i=0},s_k^{i=0}=s_{k,t}^{i=0})}\right)-\\ \nonumber
&\log\left({\max_{\{{\mathbf{\hat{s}}}_{k,t}^{i=1}\}_{t=1}^{S/2}}p(\mathbf{r}|\mathbf{\hat{s}}_{k,t}^{i=1},s_k^{i=1}=s_{k,t}^{i=1})}\right)
\end{align}
\item By doing so, we need to solve $S/2$ optimization problems for each bit. In order to obtain the gradient for \eqref{eq:min_eqv}, we need $\mathcal{O}(MK^2)$ flops. Unlike our proposed method, the complexity of max-log approximation, which is $S/2\times{S^{K-1}}$, grows with $K$ exponentially.
\end{itemize}

\subsection{Soft Decision Metric for The Uplink Including Channel Estimation Error (Coded Modulation)}
\label{sec:ch_est_err}
In this section we consider the bit detection problem for the uplink, also incorporating the channel estimation error into the analysis. Denote the channel estimation error for the $k^{th}$ user by:
\begin{equation}
\mathbf{\tilde{h}}_k=\mathbf{h}_k-\mathbf{\hat{h}}_k.
\end{equation}
The received signal in \eqref{eq:symbolmodel} can be rewritten as:
\begin{equation}
\mathbf{r}=\sum_{k=1}^K\sqrt{p_k}\mathbf{\hat{h}}_ks_k+\sum_{k=1}^K\sqrt{p_k}\mathbf{\tilde{h}}_ks_k+\mathbf{n}.
\end{equation}
Since we do not have a closed-form expression for $\mathbf{\hat{h}}_k$, the BS may calculate the variance of each realization of $\mathbf{\tilde{h}}_k$ empirically. From \eqref{eq:comp_Gaus}, the variance of a complex isotropic Gaussian distribution is 4 times  the dispersion parameter. Heuristically, all channel estimation errors are considered to have a complex isotropic Cauchy distribution. This approach is  heuristic because one cannot calculate the variance of the complex Cauchy distribution but our aim is a simple model to find additional dispersion for the noise. Let us denote the dispersion of the $k^{th}$ user's channel estimation error by $\gamma_k$. Using Proposition 1, the received signal can be expressed as:
\begin{equation}
\mathbf{r}=\sum_{k=1}^K\sqrt{p_k}\mathbf{\hat{h}}_ks_k+\mathbf{\tilde{n}},
\end{equation}
where $\mathbf{\tilde{n}}$ is assumed to be the complex isotropic Cauchy distribution with the dispersion
$\tilde{\gamma}=\sum_{k=1}^K\sqrt{p_k}\gamma_k+\gamma$. We can replace $\gamma$ by $\tilde{\gamma}$ in the conditional probability density functions defined in \ref{up_ach} and \ref{bit_det} to calculate the uplink BER and the uplink achievable rate.

\subsection{Soft Decision Metric for The Downlink (Coded Modulation)}
For the downlink part, the transmitted signal can be expressed:
\begin{equation}
\mathbf{x}=\mathbf{A}\mathbf{s}=\sum_{k=1}^K\mathbf{a}_ks_k,
\end{equation} 
where $\mathbf{A}$ is an $M\times{K}$ precoder matrix, $\mathbf{a}_k$ is the $k^{th}$ column of $\mathbf{A}$, and $\mathbf{s}$  is an ${K}\times1$ vector including the transmitted symbols. 
Notice that $\|\mathbf{a}_k\|_2=1$. The precoder matrix is designed based on the channel estimates. We consider the   MR and ZF precoders, for which:
\begin{equation}
\mathbf{A}^{\text{MR}}=\mathbf{\hat{H}}^*\mathbf{D}^{\text{MR}},
\end{equation}
\begin{equation}
\mathbf{A}^{\text{ZF}}=\mathbf{\hat{H}}^*(\mathbf{\hat{H}}^T\mathbf{\hat{H}}^*)^{-1}\mathbf{D}^{\text{ZF}},
\end{equation}
where $\mathbf{D}^{\text{MR}}$ and $\mathbf{D}^{\text{ZF}}$ are $K\times{K}$ diagonal matrices for  normalization purposes, and the $k^{th}$ column of $\mathbf{\hat{H}}$ is $\mathbf{\hat{h}}_k$.

The $k^{th}$ user receives  the following signal:
\begin{equation}
{y}_k=\sqrt{p_k^k}\mathbf{h}_k^T\mathbf{a}_ks_k+\sum_{l=1,l\neq{k}}^K\sqrt{p_l^k}\mathbf{h}_k^T\mathbf{a}_ls_l+n_k,
\end{equation}
where ${p_l^k}$ is the received power  corresponding to the symbol $s_l$  at the user $k$, and $n_k$ is the complex isotropic Cauchy noise at the $k^{th}$ user.

Note that neither the BS nor the users  know the real channels. 
Moreover, the users do not have access to any channel estimates. 
Therefore, for the downlink decoding we assume that the $k^{th}$ user only knows the corresponding channel gain, which is given by:
\begin{equation}
g_k^{\text{MR}}=\|\mathbf{\hat{h}}_k\|_2,
\end{equation}
\begin{equation}
g_k^{\text{ZF}}=\mathbf{D}^{\text{ZF}}_{kk},
\end{equation}
where $\mathbf{D}^{\text{ZF}}_{kk}$ refers to the $k^{th}$ diagonal element. 
The LLR expression for the $i^{th}$ bit of the $k^{th}$ symbol when the MR precoder is used can be expressed as:
\begin{equation}
\label{LLR_downlink}
l_{k,i}^{\text{MR}}=\log\left(\frac{p(b_{k,i}=0|{y_k})^{\text{MR}}}{p(b_{k,i}=1|{y_k})^{\text{MR}}}\right)=\log\left(\frac{\displaystyle\sum_{{s}:b_{i}({s})=0}p({y}_k|{s})^{\text{MR}}}{\displaystyle\sum_{{s}:b_{i}({s})=1}p(y_k|s)^{\text{MR}}}\right),
\end{equation}
where $p(y_k|s)^{\text{MR}}$ is given by:
\begin{equation}
\label{prob_MR}
p(y_k|s)^{\text{MR}}=\frac{\gamma}{2\pi\left(\gamma^2+\left|y_k-\sqrt{p_k^k}g_k^{\text{MR}}s\right|^2\right)^{3/2}}.
\end{equation}
To find $p(y_k|s)^{\text{ZF}}$, one can replace $g_k^{\text{MR}}$ by $g_k^{\text{ZF}}$.

\section{Simulation Results}
\label{sec:Sim_res}

In this section, we present simulation results. First, we compare the performances of the channel estimates in terms of uncoded SER. Then we present  achievable rate results for the perfect and imperfect CSI cases. Finally we present coded BER curves and compare their behavior
with the predictions from the achievable rate analysis. 

\subsection{Performances of  the Channel Estimates}
\begin{figure}[h]
	\centering
	\includegraphics[width=\columnwidth]{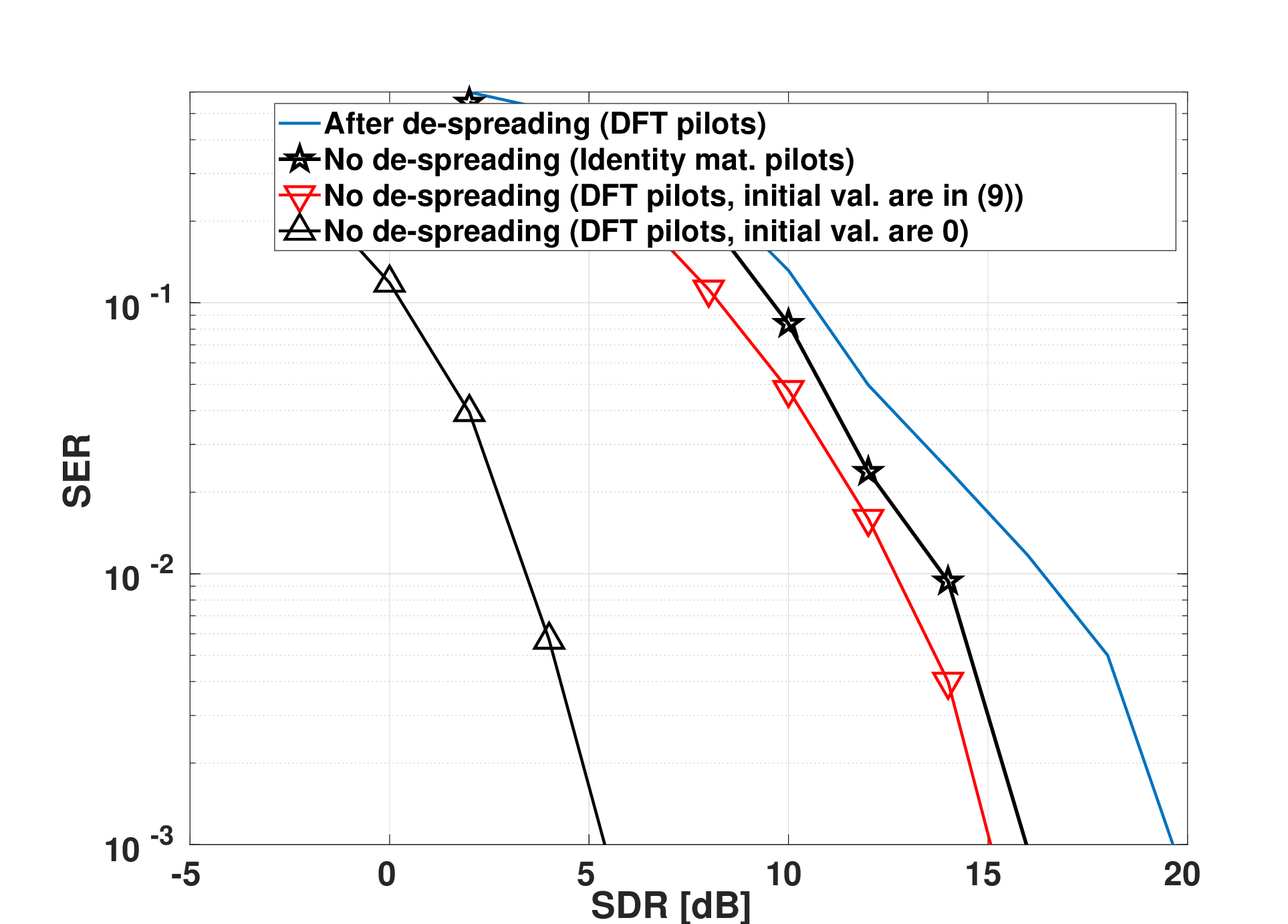}
	\caption{Comparison of the performance of different channel estimation approaches,
	quantified in terms of (uncoded) SER.}
	\label{SER}
\end{figure}

The simulation parameters are as follows: The number of antennas is $M=100$. 
The number of users and the pilot length are $K=8$ and $\tau=15$, respectively. 
The pilot signals are chosen from the normalized DFT matrix. 
The channel matrix contains realizations of circularly symmetric Gaussian random variables with unit variance. 
The dispersion parameter of the Cauchy noise is normalized to unity. 
We fix the received signal powers of 7 users such that these powers range from 1 to 7 dB. 
We change the received signal power of the remaining  user  and  observe the effect on the performance. 
The parameters of the simulation  are summarized in Table \ref{tab:th}.

\begin{table} [h]
    \centering
        \caption{Parameters for the simulations }
          \label{tab:th}
    \begin{tabular}{|c|c|}
  
\hline
               $M$ (number of antennas)& 100   \\\hline
               $K$ (number of users)   & 8 \\\hline
 $\tau$ (pilot length)   & 15 
\\\hline
$\gamma$ (dispersion parameter) & 1 \\\hline
 The received signal powers of 7 users [dB]& $(1:1:7)$  
\\\hline
$T$ for Fig. \ref{SER} and Fig. \ref{mismatched} (length of the coherence block)& $215$  
\\\hline
$T$ for all other figures (length of the coherence block)& $339$  
\\\hline
Number of coherence blocks simulated for each SDR& $500$  
\\\hline

    \end{tabular}
\end{table}

\begin{figure}[h]
	\centering
	\includegraphics[width=\columnwidth]{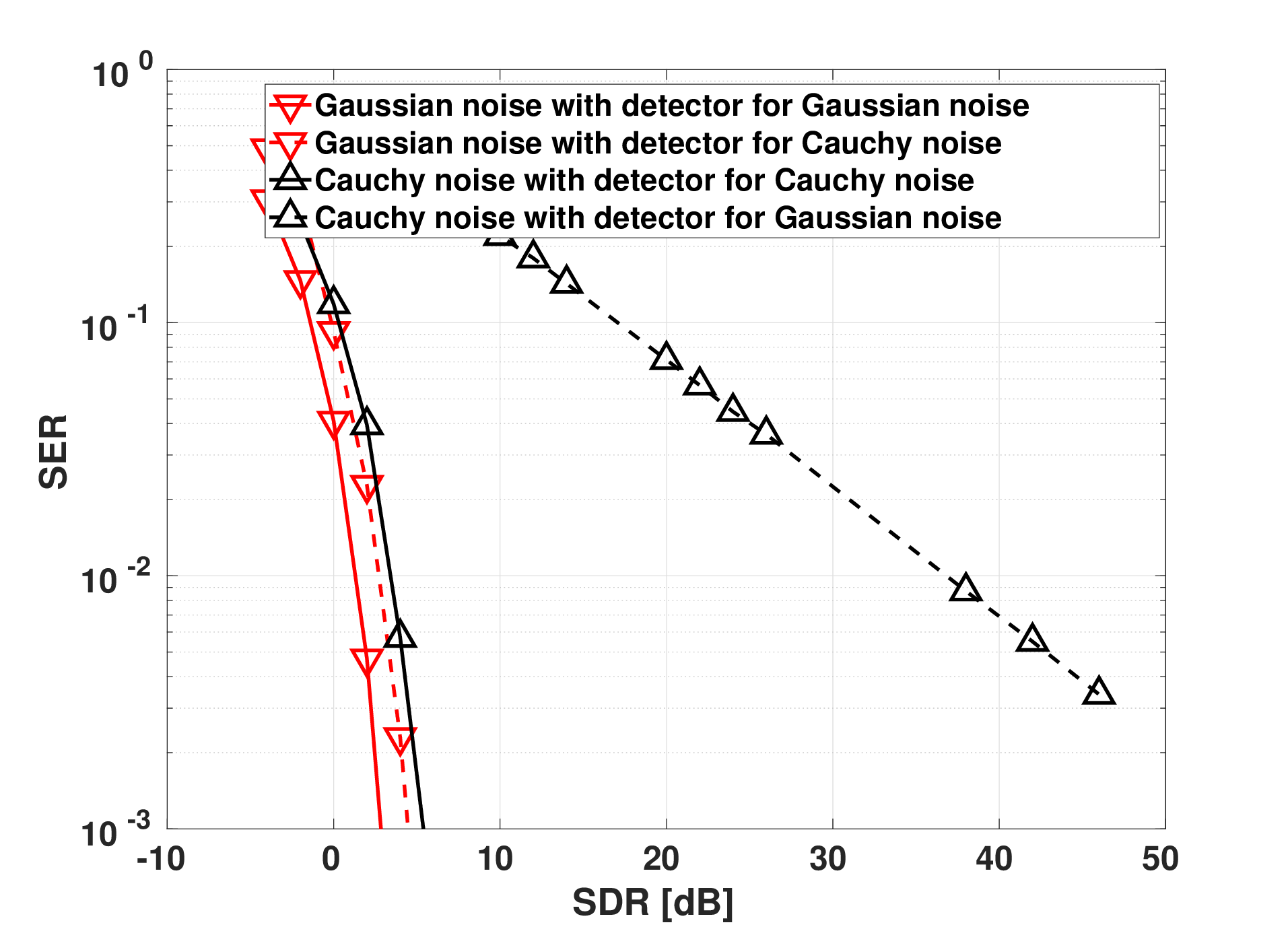}
	\caption{Performances of the detectors in the presence of Cauchy and Gaussian noises,
	quantified in terms of (uncoded) SER.}
	\label{mismatched}
\end{figure}

We present the effects of the channel estimates on the symbol detection. We have three types of channel estimates: the estimates obtained from the signal after de-spreading, the estimates obtained from the unprocessed signal where the initial values for the algorithm in Section \ref{sec:est_unprocess} are taken from \eqref{eq:ML}, and the estimates obtained from the unprocessed signal where the initial values for the algorithm in Section \ref{sec:est_unprocess} are zero. 
To generate symbols, each user transmits 200 quadrature-phase-shift keying (QPSK) symbols for each coherence block, whose length is taken to be $T=215$. 
The small-scale fading for each channel vector is created 500 times so we have $10^5$ symbols for each SDR.

 We present SER performances in Fig. \ref{SER}. 
 Based on Fig. \ref{SER}, the detector using the channel estimates obtained from the unprocessed signal outperforms   the detector using  the channel estimates obtained via the de-spreading operation. One important observation is that the choice of pilot books is important even if the pilot books are chosen from any unitary matrix. For example, based on  Fig. \ref{SER}, if the pilots are chosen from the normalized DFT matrix, we have much better SER performance than the case where the pilots are chosen from the identity matrix. This situation does not appear in the Gaussian noise because the pilot books chosen from any unitary matrix give the same performances \cite[Chapter~3]{marzetta_larsson_yang_ngo_2016}. Therefore, one can conclude that if the noise is Cauchy, one can split the power of the pilots and can increase the number of signal samples in the receiver side.
 Another important observation is that the channel estimates obtained from the unprocessed signal are sensitive to the initial values used in the algorithm in Section \ref{sec:est_unprocess}; this is because the likelihood function of the Cauchy distribution is neither log-concave nor log-convex so it has has many local minima. Based on this,  $\hat{\mathbf{h}}_k^\text{ML}$ includes outliers causing us to get trapped at poor local optima. Since the mean of the channel realizations is zero, it is expected to obtain a better local optimum point when the initial values are zero.
 
 Quantitatively, when the SER is $10^{-3}$, the required SDR for the best detector is almost 5 dB. For SER $10^{-3}$, the performance gaps to the second and the third detectors compared with the best detector are almost 10 and 15 dB, respectively.

In Fig. \ref{mismatched} we present the performances of two detectors in the presence of two types of noise: Gaussian noise-detector for Gaussian noise, Gaussian noise-detector for Cauchy noise, Cauchy noise-detector for Cauchy noise, and Cauchy noise-detector for Gaussian noise. Note that we change the noise only in the data phase and all detectors use the best channel estimates that are obtained under the Cauchy noise. Based on Fig. \ref{mismatched}, the performance gap between the detector for Cauchy and the detector for Gaussian is  small in the presence of  Gaussian noise. On the other hand, the performance of the detector designed for Gaussian noise is quite poor when the noise is Cauchy. The conclusion is that the detector for Gaussian noise is not robust against the outliers.

For the rest of the paper, we only use the best channel estimates.

\subsection{Achievable rates with the Cauchy Noise}

In this section, we first present the uplink achievable rates of the communication link for both the perfect and  imperfect CSI cases. 
For the modulation scheme, we use QPSK again. The BS has either 100 or 4 antennas and serves a single user. 
For the imperfect CSI case, the length of the coherence block is $T=339$ (we will explain why we choose 339 later). 
 
\begin{figure}[h]
	\centering
	\includegraphics[width=\columnwidth]{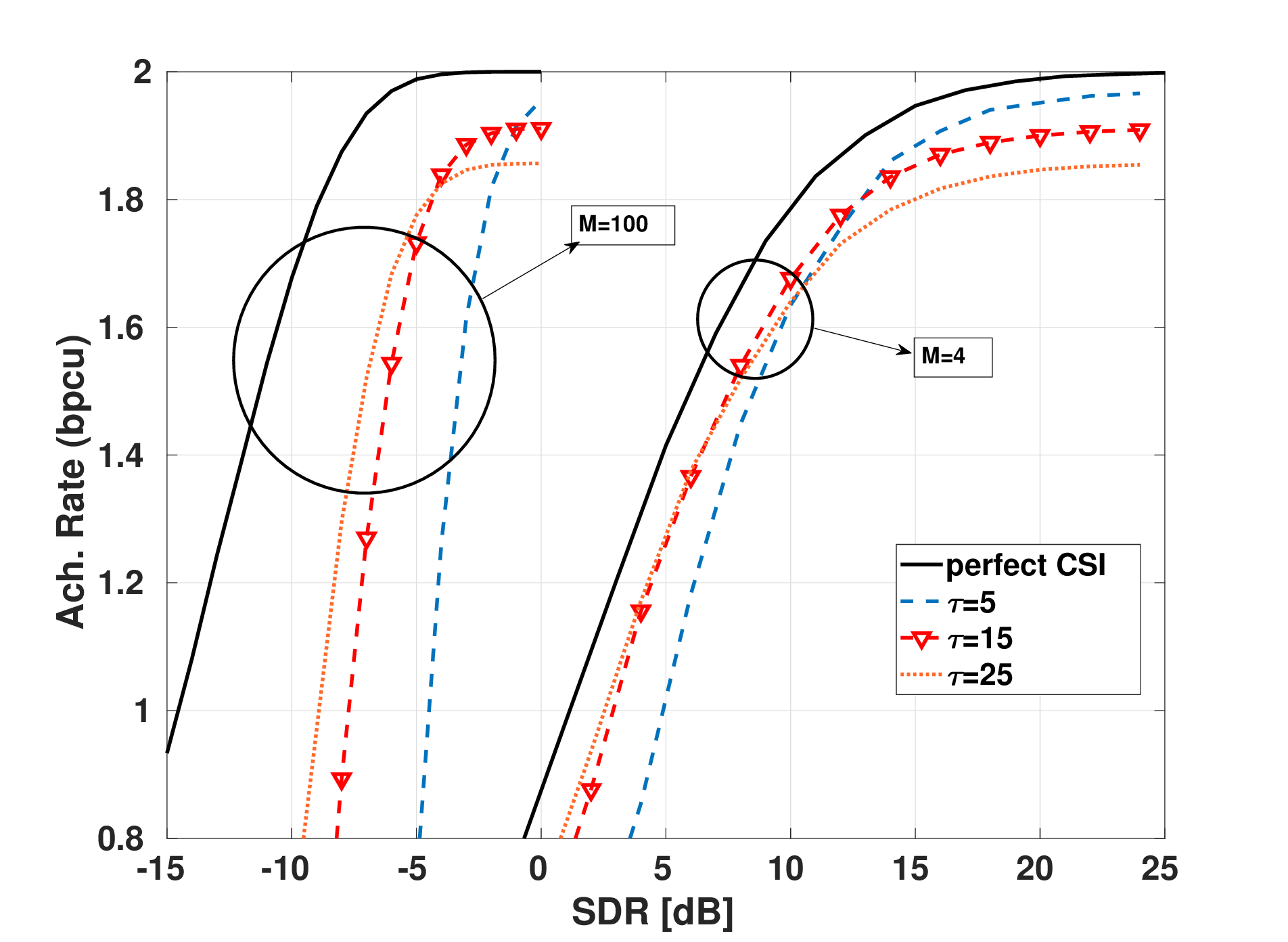}
	\caption{Uplink achievable rates  for the perfect and imperfect CSI cases. 
	Here  $\gamma_k$ in the detector is not adjusted to take into account the presence of channel estimation errors. }
	\label{uplink}
\end{figure}

\begin{figure}[h]
	\centering
	\includegraphics[width=\columnwidth]{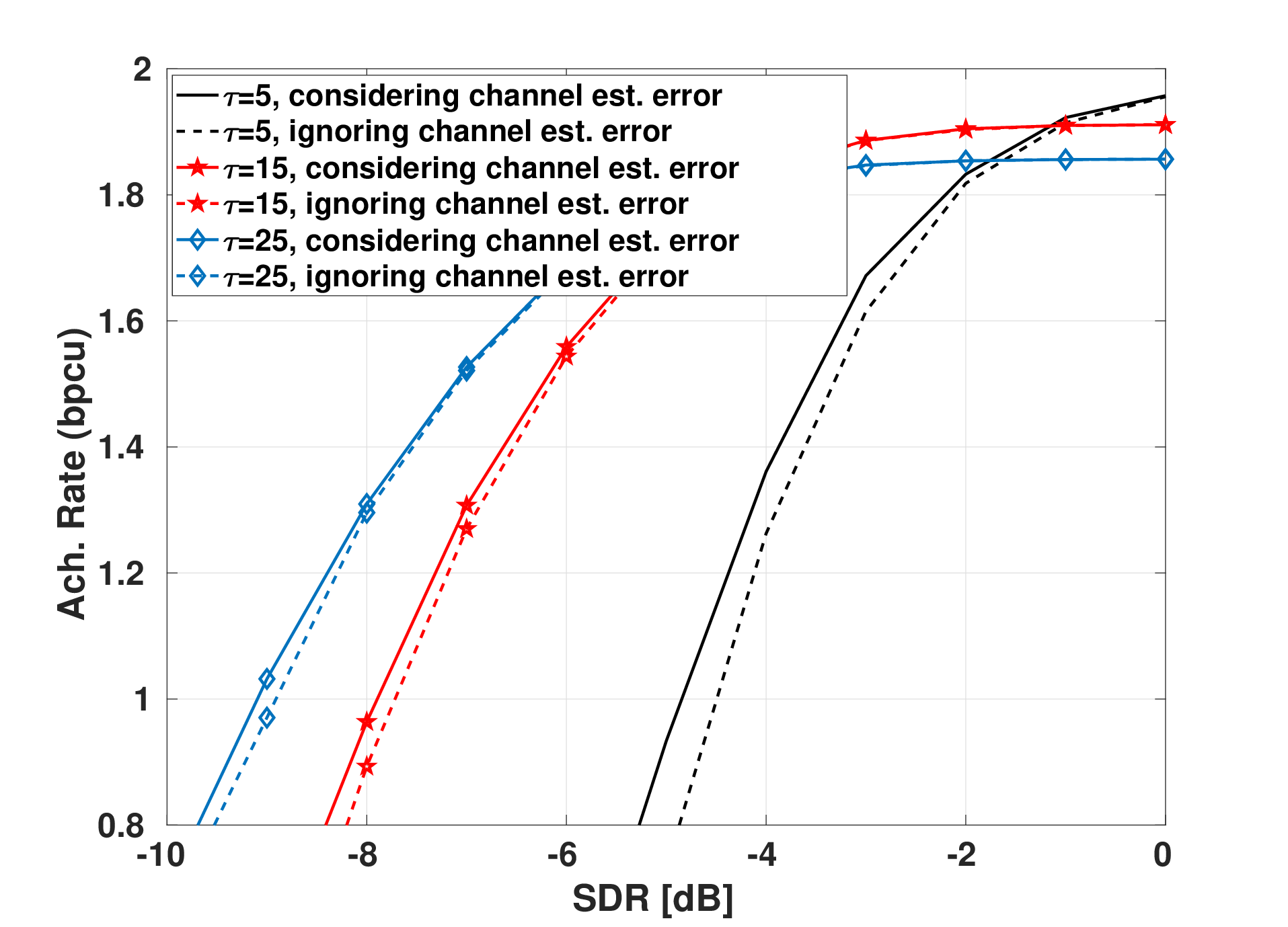}
	\caption{Comparison of uplink achievable rates ($M=100$)  with (``considering'') and without (``ignoring'') adjustment of $\gamma_k$  to take into account
	the presence of channel estimation errors, as described in Section~\ref{sec:ch_est_err}.}
	\label{uplink_ch_error}
\end{figure}

\begin{table*} [h]
    \centering
        \caption{The uplink BER performances (Fig. \ref{BER_uplink} )}
          \label{tab:uplink}
    \begin{tabular}{|c|c|c|c|}

\hline
              Uplink& The Bounds for Ach. Rate [dB] & Decoding Thr. [dB] & Offset [dB]  \\\hline
             $M=100$, $K=1$ & -6.4 & -5.5& 0.9 \\\hline
            
             $M=100$, $K=2$ & -6.4 & -4.3& 2.1 \\\hline
             $M=100$, $K=8$ & -6.4 & 1.3& 7.7 \\\hline
             $M=4$, $K=1$ &  6.7 & 9.9& 3.2 \\\hline

    \end{tabular}
\end{table*}

In Fig. \ref{uplink}, we present uplink performances of the communication link. From Fig. \ref{uplink}, we observe that when we increase the number of antennas we can obtain the same achievable rates with less SDR. For example, when $\tau$ and the desired rate are chosen to be 15 and 1.8 bits-per-channel-use (bpcu), the required SDR for $M=100$ is almost 20 dB less than the required SDR for $M=4$. Another observation is that when we increase the pilot length, we obtain better performances  in low SDR. 
However in high SDR cases, we do not need to use long pilot sequences. 
Also in any case, we cannot obtain the same performance as in  the perfect CSI case, because of the pre-log factors appearing in the achievable rate. 
In Fig. \ref{uplink_ch_error}, we present the uplink achievable rate curves for the two cases that the BS considers the channel estimation error, and ignores it, respectively. 
We observe that when $\tau$ increases the gap between two achievable rate curves, i.e., the curve obtained when the BS ignores the channel estimation error and the curve obtained when the BS considers the channel estimation error, gets closer to the each other. This result is expected because when  $\tau$ increases, the channel estimation error decreases.

\begin{figure}[h]
	\centering
	\includegraphics[width=\columnwidth]{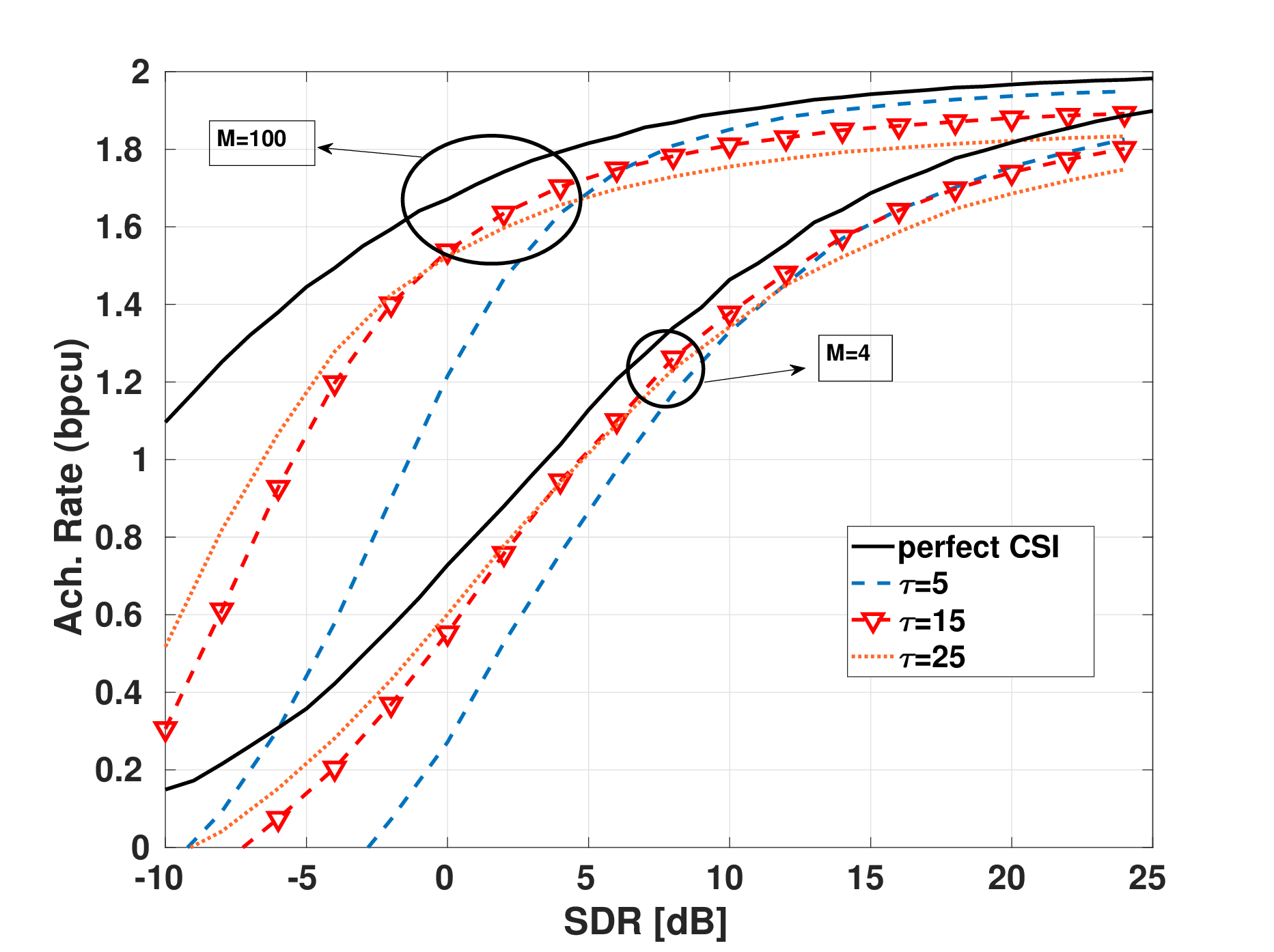}
	\caption{Downlink achievable rates in the perfect and imperfect CSI cases.}
	\label{downlink}
\end{figure}

Next,  in Fig. \ref{downlink}, we present the downlink achievable rates. Again for the downlink, when we increase the number of antennas, we need less SDR for the same achievable rates and for the low SDR cases we may need longer pilot sequences. Another interesting observation is that the uplink-downlink duality does not hold for the Cauchy noise even for  perfect CSI, 
in contrast to  the case of  Gaussian noise. With Gaussian noise, when the MR decoder/precoder is used, the structure of the signal is the same both on uplink and downlink. However, this is not the case for the Cauchy noise.

From Figs. \ref{uplink} and   \ref{downlink}, we do not observe uplink-downlink duality which is consistent with  Remark~4 in Section~\ref{sec:down}.

\begin{figure}[h]
	\centering
	\includegraphics[width=\columnwidth]{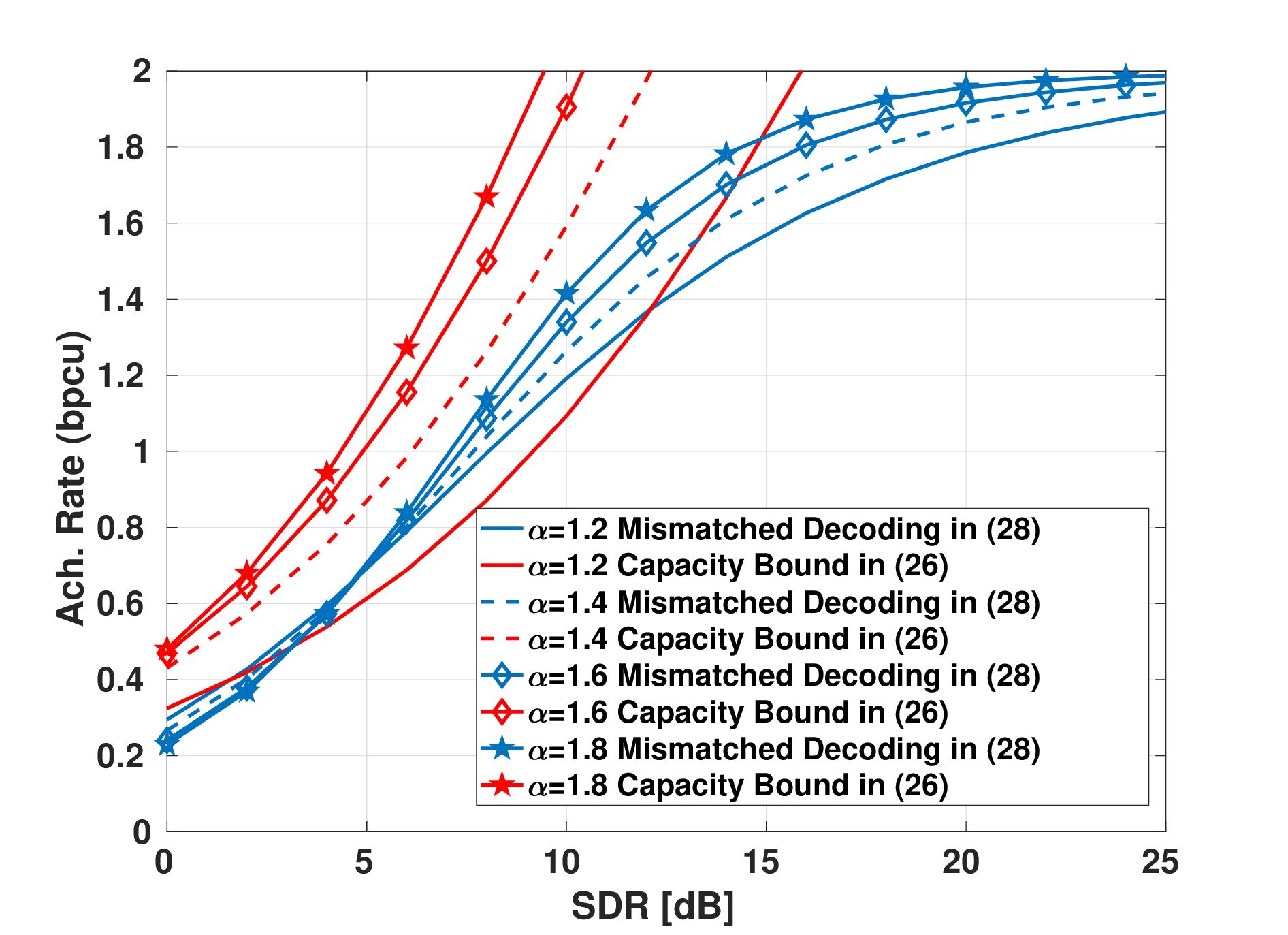}
	\caption{S($1<\alpha<2$)S noise vs. Cauchy Decoder: Capacity bounds in \eqref{eq:Cap_bound} and mismatched achievable rate for QPSK in \eqref{eq:mis_rate}   with different S$\alpha$S noise.}
	\label{mismatched_cauchy}
\end{figure}

In Fig. \ref{mismatched_cauchy}, we 
present the capacity bound in \eqref{eq:Cap_bound} and the mismatched achievable rate in \eqref{eq:mis_rate} for QPSK modulation. We generate S$\alpha$S noise with unit dispersion for different $\alpha$. From Fig. \ref{mismatched_cauchy}, if the noise becomes more impulsive, the gap between the capacity bound and the mismatched achievable rate decreases. Let us focus on code rate $3/4$, which corresponds to 1.5 bpcu for QPSK modulation. Numerically, the gaps between these two metrics in \eqref{eq:Cap_bound} and \eqref{eq:mis_rate} become $3.7, 3.5, 3.1$ and 0.9 dB with $\alpha=1.8, \alpha=1.6, \alpha=1.4$ and $\alpha=1.2$, respectively. In \cite{7866884}, it is shown   that the capacity bound becomes tighter for smaller $\alpha$. In conclusion, Fig. \ref{mismatched_cauchy} demonstrates that a decoder metric based on the   Cauchy model can perform well in the presence of noise with any S($1<\alpha<2$)S distribution. 

\subsection{Coded BER Comparisons}

In this section, we present coded BER performance for both the uplink and the downlink. 
To do this,  QPSK modulation is used. 
For the channel coding, we use a low-density-parity-check  (LDPC) code. 
The coding rate is chosen to be  3/4 and there are 648 bits per packet after the encoder. 
Therefore, the number of transmitted symbols in each packet is 324. 
We split each packet into 9 sub-packets; each sub-packet occupies  36 symbols in a coherence block
and all sub-packets go into different coherence blocks and see different channel realizations. 
This way, a comparison with the ergodic achievable rates derived in Section~\ref{sec:Ach_rate} is justified.
The length of the pilot vector is 15 in all simulations. 
We use a coherence block length of $339$, which is reasonable for an urban area with some mobility \cite{marzetta_larsson_yang_ngo_2016}. 

Note that we divide each coherence block into either a pilot-phase plus an uplink-phase, or into a pilot-phase plus a downlink-phase. For the decoding, we first calculate the LLR values for each bit. These values are then given to the LDPC decoder, which uses belief propagation with 50 iterations.  

\begin{figure}[h]
	\centering
	\includegraphics[width=\columnwidth]{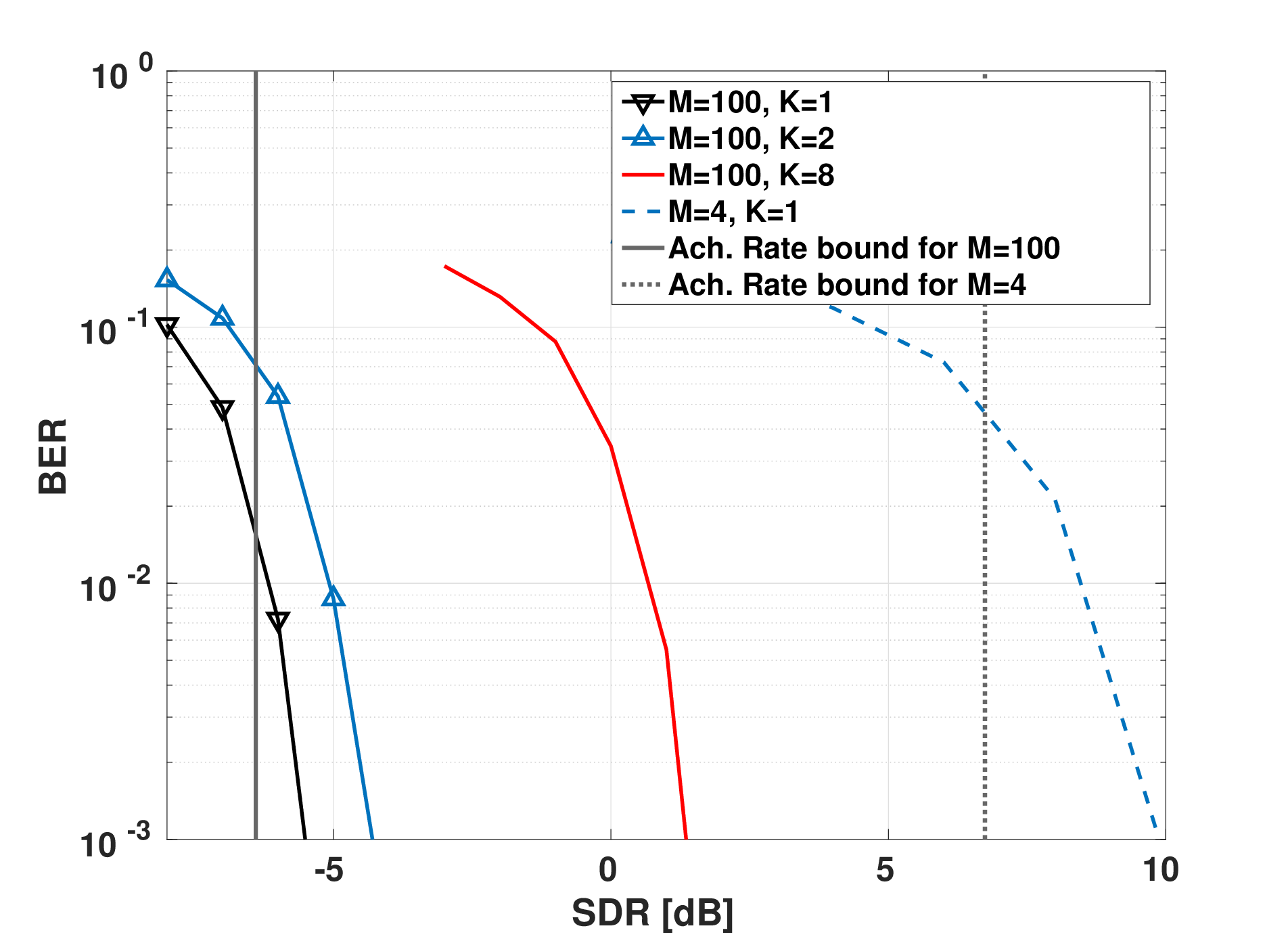}
	\caption{Empirical coded BER  for the uplink, 
 without adjusting $\gamma_k$ in the detector to take into account the presence of
   channel estimation errors.}
	\label{BER_uplink}
\end{figure}

\begin{table*} [h]
    \centering
        \caption{The downlink BER performances (Fig. \ref{BER_downlink} )}
          \label{tab:downlink}
    \begin{tabular}{|c|c|c|c|}
  
\hline
              Downlink& The Bounds for Ach. Rate [dB] & Decoding Thr. [dB] & Offset [dB]  \\\hline
             $M=100$, $K=1$  & -1.6 & 2.2& 3.8 \\\hline
            
             $M=100$, $K=2$ ZF & -1.6 & 2.4& 4 \\\hline
             $M=100$, $K=2$ MR & -1.6 & 2.6& 4.2 \\\hline
             $M=100$, $K=8$ ZF& -1.6 & 5& 6.6\\\hline
             $M=100$, $K=8$ MR& -1.6 & 5.7& 7.3\\\hline
             $M=4$, $K=1$ &  11 & 15.9& 4.9\\\hline

    \end{tabular}
\end{table*}

First, we consider the uplink performance.   Fig. \ref{BER_uplink} shows the results. 
For massive MIMO we consider  $K=1$, $K=2$ and $K=8$ users.
We also present the performance of communication link including a single user ($K=1$) and a BS ($M=4$). 
For the BER curves, the decoding threshold is defined as the SDR value at which the waterfall starts (the SDR value when BER is $10^{-3}$). The quantitative results are presented in  Table \ref{tab:uplink}. From  Fig. \ref{BER_uplink} and  Table \ref{tab:uplink}, it is clear that the BER performance gets closer to the achievable rate bound when the number of antennas increases. Note that there is a gap between the decoding threshold obtained from the BER simulations and the bound obtained by the achievable rate analysis. The gap of   massive MIMO, ($M=100, K=8$), is greater than that of the network, ($M=4, K=1$). However, the main advantage is that massive MIMO serves more than one user.

\begin{figure}[h]
	\centering
	\includegraphics[width=\columnwidth]{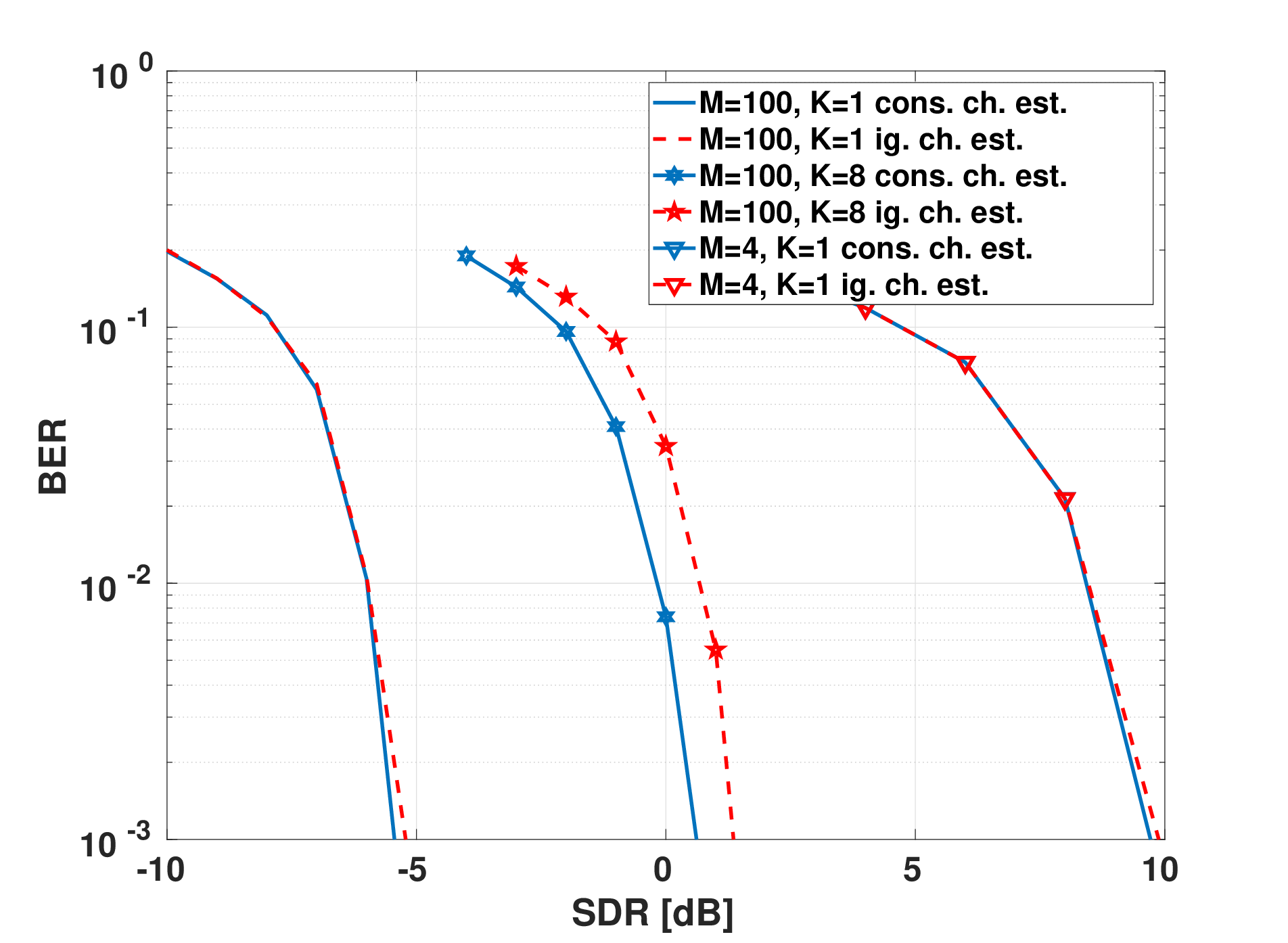}
	\caption{Comparison of coded BER, with (``cons.'') and without (``ig.'') adjustment of $\gamma_k$ for the
	presence of channel estimation errors. }
	\label{uplink_ch_error_BER}
\end{figure}

In Fig. \ref{uplink_ch_error_BER}, we present BER curves for the uplink when the BS considers the channel estimation error in decoding metric. We observe that when the number of users increases, the BER performance  improves if the BS considers the channel estimation error. Quantitatively, when the number of users is 8, the BER performance is improved by 0.8 dB. On the other hand  when the number of user is 1, this improvement becomes 0.2 dB.
\begin{figure}[h]
	\centering
	\includegraphics[width=\columnwidth]{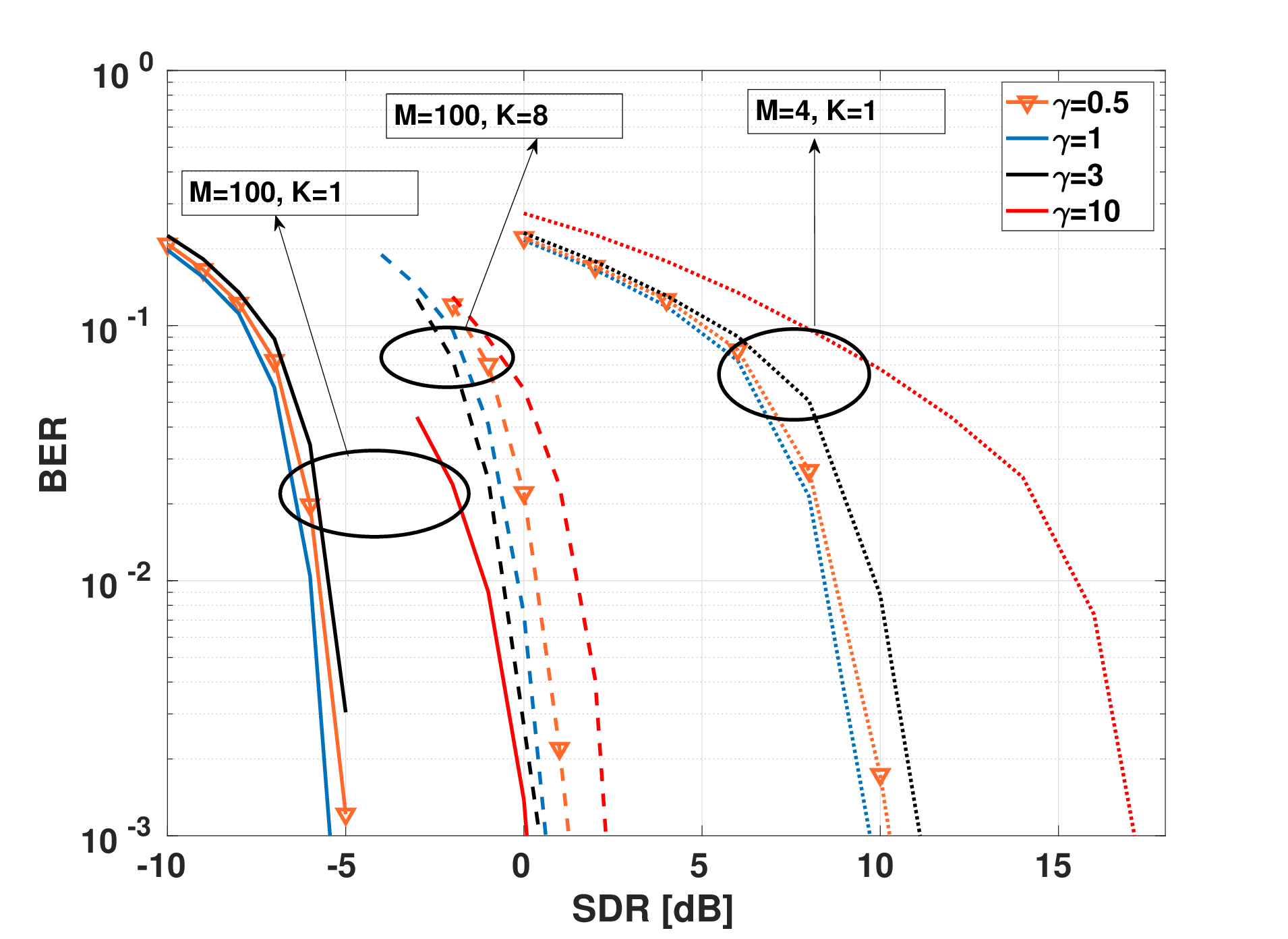}
	\caption{Coded BER curves when the dispersion used in the likelihood function is mismatched to actual real dispersion (the actual dispersion is always 1). All receivers take into account the channel estimation errors.}
	\label{uplink_mismatched}
\end{figure}

In Fig. \ref{uplink_mismatched}, we present the uplink BER curves when the noise dispersions in the likelihood functions are 0.5, 1, 3 and 10 and the dispersion of the noise   is 1. 
When the dispersion in the likelihood function is 0.5 and there is only a single user, we have a relatively low performance loss:  around 0.7 dB when $M=100$. 
Moreover, when $K=8$ the receiver whose dispersion is 3 outperforms the receiver whose dispersion is 1 because of  interference.

\begin{figure}[h]
	\centering
	\includegraphics[width=\columnwidth]{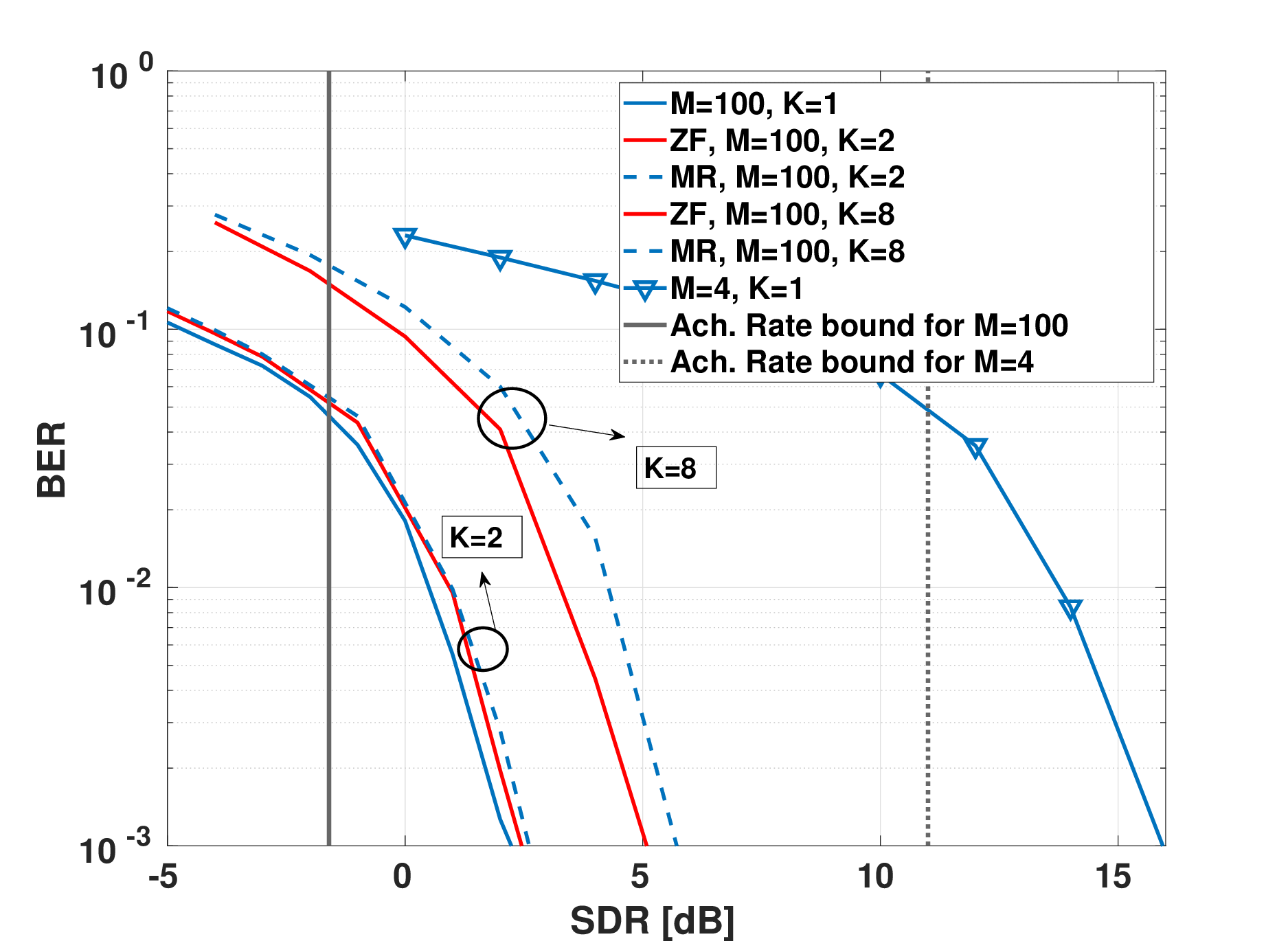}
	\caption{Empirical coded BER  for the downlink.}
	\label{BER_downlink}
\end{figure}

Similarly, we obtained  BER performances for the downlink; see  Fig. \ref{BER_downlink}. 
Compared with the uplink of massive MIMO, we have extra curves corresponding to the two different  precoders,  MR and ZF. 
Like for the uplink, in the downlink having more antennas implies that the performance gets closer to the   bounds obtained from the achievable rate analysis.

Another observation is that the ZF precoder performs slightly better than the MR precoder when the number of users is 8. 
These precoders perform almost the same when the number of users is 2. 
This is expected because the user-interference is so small when there are 2 users so these precoders are almost identical. The numerical results are presented in Table \ref{tab:downlink}.

\section{Conclusions}

We  investigated massive MIMO with  Cauchy noise from three perspectives:  channel estimation, achievable rates and soft bit detection. First, we obtained the channel estimates from  uplink pilots in two ways: with and without  de-spreading the received pilot signal. In contrast to the
case of Gaussian noise, with Cauchy noise the de-spreading operation does not result in a
sufficient statistic for the channel estimation. Consequently, better channel estimation performance is  obtained when using  the unprocessed received signal. 

Next, we obtained uplink and downlink achievable rates for the cases of perfect and imperfect CSI (using the channel estimates we developed). 
We observed that for low SDR, longer pilot signals can be used to obtain better achievable rates. Also, the achievable rate increases with an increasing number of antennas on both uplink and downlink. We also compared the performance of Cauchy decoding with a capacity bound for general S$\alpha$S noise  available in the literature. Through simulations, we validated the technical soundness of using   the Cauchy receiver for other S$\alpha$S noise.

We compared the  detectors designed for the Cauchy and Gaussian noises in the presence of both types of noises. An important observation is that the performance losses both of  the detector designed for Cauchy noise and of the detector designed for Gaussian noise are small in the presence of Gaussian noise. However, the detector designed for Gaussian noise works poorly in the presence of Cauchy noise. This means that unlike the Gaussian-noise detector, the Cauchy-noise detector is very robust, and should be a preferred choice whenever robustness to unknown noise distributions is a priority.

Finally, we obtained metrics for soft bit detection. Based on these metrics, we performed
numerical simulations of the BER, using  LDPC coding and QPSK modulation. We compared the threshold of this  BER performance with the achievable rate bound. The main conclusions are that the gap between the decoding threshold and achievable rate bound is small, and that this gap decreases with an increasing number of antennas.

\appendices
\section{Mathematical Preliminaries}
\label{sec:Math_pr}
A real-valued random variable S$\alpha$S is defined by its characteristic function \cite{Niki}:
\begin{equation}
\phi(t)=\exp(j\delta{t}-\gamma|t|^{\alpha}),
\end{equation}
where $t\in\mathbb{R}$, $\delta$ is the location parameter, $\gamma$ is the dispersion parameter which determines the spread of the distribution in $\gamma>0$, and $\alpha$ is the characteristic exponent which
satisfies  $0<\alpha\leq{2}$ and determines the tail level of distribution. 
A smaller $\alpha$ yields a more impulsive and heavily-tailed distribution, and vice versa. 
There are two important special cases of the S$\alpha$S distribution:  Cauchy ($\alpha=1$) and Gaussian   ($\alpha=2$), which both have pdfs in  closed form.
These pdfs, if  $\delta=0$, are given by 
\begin{equation}
\label{Cauch}
c(x)=\frac{\gamma}{\pi(x^2+\gamma^2)},
\end{equation}
for the Cauchy case and 
\begin{equation}
\label{Gauss}
g(x)=\frac{1}{\sqrt{4\pi\gamma}}\exp\left(-\frac{x^2}{4\gamma}\right),
\end{equation}
  for the Gaussian. 
From \eqref{Cauch} and \eqref{Gauss}, two important results can be obtained: For the Gaussian distribution, the variance is $2\gamma$. For the Cauchy distribution, the mean and variance  are undefined. More precisely the variance is infinite because $\mathbb{E}[\mathbf{|X|}^p]=\infty$ if $p\geq\alpha$.

In communication problems, we generally deal with complex-valued random variables.
Define the complex Cauchy random variable $X=X^R+jX^I$ where $X^R$ and $X^I$ are jointly S($\alpha=1$)S. By definition, the characteristic function is:
\begin{equation}
\label{eq:char}
\phi_X(\omega)=\mathbb{E}\left[\exp(j\Re{(\omega{X}^*)})\right],
\end{equation}
where $\omega\in\mathbb{C}$. $X$ is an isotropic complex Cauchy random variable if and only if the characteristic function has the form \cite[Chapter~3]{Panagi}:
\begin{equation}
\phi_X(\omega)=\exp(-\gamma|w|).
\end{equation} 
The pdf of an isotropic complex Cauchy distribution is:
\begin{equation}
\label{eq:Joint_cauchy}
f_{X}(x)=\frac{\gamma}{2\pi(|x|^2+\gamma^2)^{3/2}}.
\end{equation}
As the  name ``isotropic'' suggests, the pdf is invariant to a rotation of the complex angle and depends  only on the magnitude of the   realization from \eqref{eq:Joint_cauchy}. 
The marginal distributions of $X^R$ and $X^I$ can be easily obtained and they are same as for the pdf in \eqref{Cauch}. 
An important property of the isotropic  complex Cauchy distribution is that $X^R$ and $X^I$ are  statistically dependent. 
This can be immediately seen from the fact that the product of the pdfs of $X^R$ and $X^I$  is not equal to the joint pdf in \eqref{eq:Joint_cauchy}. 
This is a fundamental difference between the isotropic  complex Cauchy and isotropic  complex Gaussian distributions. (For the latter,   in-phase and quadrature components are independent, which is also
referred to as  circular symmetry of the noise  \cite[Chapter~3]{gallager2013stochastic}.) 
For the sake of completeness, the pdf of an isotropic complex Gaussian  random variable, $Y$, (which necessarily has   zero mean) is given by:
\begin{equation}
\label{eq:comp_Gaus}
f_{Y}(x)=\frac{1}{4\pi\gamma}\exp\left(-\frac{|x|^2}{4\gamma}\right).
\end{equation}

\section{Solving The Optimization Problem in \eqref{obj}}
\label{sec:opt_pro}
We denote the objective function in \eqref{obj} by  $f$. We use the gradient descent algorithm \cite[Chapter~3]{nocedal2006numerical}. 
The optimization variables are $\mathbf{h}_1^R[1]$ and $\mathbf{h}_1^{I}[1]$, corresponding to the real and imaginary parts of $\mathbf{h}_1[1]$. 
Before defining the gradient vector, let us define the following  auxiliary functions:
\begin{align}
p(\mathbf{h}_1^R[1],\mathbf{h}_1^{I}[1],&i)= \\ \nonumber
&\mathbf{Y'}^R[1,i]-\sqrt{\tau{p}_1}\left(\mathbf{h}_1^R[1]\boldsymbol{\phi}_1^R[i]-\mathbf{h}_1^I[1]\boldsymbol{\phi}_1^I[i]\right),
\end{align}
\begin{align}
r(\mathbf{h}_1^R[1],\mathbf{h}_1^{I}[1],&i)= \\ \nonumber
&\mathbf{Y'}^I[1,i]-\sqrt{\tau{p}_1}\left(\mathbf{h}_1^I[1]\boldsymbol{\phi}_1^R[i]+\mathbf{h}_1^R[1]\boldsymbol{\phi}_1^I[i]\right).
\end{align}
The gradient of the objective function in \eqref{obj} can now be expressed as:
\begin{align}
&\nabla{f}\begin{bmatrix}\mathbf{h}_1^R[1]\\ \mathbf{h}_1^I[1]\end{bmatrix} = -2\sqrt{\tau{p}_1} \times \\ \nonumber
&\begin{bmatrix}
\displaystyle\sum_{i=1}^\tau\frac{\boldsymbol{\phi}^R_1[i]p(\mathbf{h}_1^R[1],\mathbf{h}_1^{I}[1],i)+\boldsymbol{\phi}^I_1[i]r(\mathbf{h}_1^R[1],\mathbf{h}_1^{I}[1],i)}{\gamma^2+|\mathbf{Y'}[1,i]-\sqrt{\tau{p}_1}\mathbf{h}_1[1]\boldsymbol{\phi}_1[i]|^2}\\ 
\displaystyle\sum_{i=1}^\tau\frac{-\boldsymbol{\phi}^I_1[i]p(\mathbf{h}_1^R[1],\mathbf{h}_1^{I}[1],i)+\boldsymbol{\phi}^R_1[i]r(\mathbf{h}_1^R[1],\mathbf{h}_1^{I}[1],i)}{\gamma^2+|\mathbf{Y'}[1,i]-\sqrt{\tau{p}_1}\mathbf{h}_1[1]\boldsymbol{\phi}_1[i]|^2}
\end{bmatrix}.
\end{align}
With the  gradient descent approach, we update the channel vector according to:
\begin{equation}
\begin{bmatrix}\mathbf{h}_1^R[1]\\ \mathbf{h}_1^I[1]\end{bmatrix}^{j+1}=\begin{bmatrix}\mathbf{h}_1^R[1]\\ \mathbf{h}_1^I[1]\end{bmatrix}^j-\eta_j\nabla{f}\left(\begin{bmatrix}\mathbf{h}_1^R[1]\\ \mathbf{h}_1^I[1]\end{bmatrix}^j\right).
\end{equation}
where $\begin{bmatrix}\mathbf{h}_1^R[1]\\ \mathbf{h}_1^I[1]\end{bmatrix}^j$ and $\eta_j$ are the vectors that represent the solutions  and the step length for the  $j^{th}$ iteration, respectively. $\eta_j$ can be chosen differently for each iteration using, for example,  the backtracking algorithm \cite[Chapter~3]{nocedal2006numerical}. Note that the cost of obtaining the gradient for one channel realization is $\mathcal{O}(\tau)$ flops, and for all channel realizations the cost is $\mathcal{O}(MK\tau)$ flops.

\section{Empirical Calculation of SISO Channel's Mutual Information}
\label{sec:mut_inf}
The mutual information between  channel input $X$, and channel output $Y$ can be defined as:
\begin{equation}
\label{eq:mut}
I(X;Y)=H(X)-H(X|Y).
\end{equation}
Let us assume that $X$ is chosen uniformly from a certain discrete set $A_X$ with cardinality $S$. Then, \eqref{eq:mut} can be written as:
\begin{equation}
I(X;Y)=\log_2{S}-\int\sum_{x\in{A_X}}-p(x,y)\log_2(p(x|y))dy,
\end{equation}
where $p(x,y)$ is the joint probability density function of $X$ and $Y$, and  $p(x|y)$ is the conditional probability density function. The second term is the expectation of $(-\log_2(p(x|y)))$ with respect to $X$ and $Y$. By using   Bayes' rule, we can re-write mutual expression as:
\begin{align}
\label{eq:Mut_inf}
I(X;Y)& =\log_2{S}-\mathbb{E}_{X,Y}\left\{\log_2\frac{p(Y)}{p(X,Y)}\right\}, \\ \nonumber
& =\log_2{S}-\mathbb{E}_{X,Y}\left\{\log_2\frac{\sum_{x\in{A_X}}p(Y|x)p(x)}{p(Y|X)p(X)}\right\}, \\ \nonumber
& =\log_2{S}-\mathbb{E}_{X,Y}\left\{\log_2\frac{\sum_{x\in{A_X}}p(Y|x)}{p(Y|X)}\right\}. \nonumber
\end{align}

Now, we consider a perfect CSI case where the receiver has the perfect knowledge of channel. The mutual information is given \cite{841172}:
\begin{equation}
\label{eq:CSI}
I(X;Y|H)=H(X|H)-H(X|H,Y).
\end{equation}
Note that
$X$ is independent of $H$. By following the steps in \eqref{eq:Mut_inf}, \eqref{eq:CSI} can be rewritten as:
\begin{equation}
\label{eq:perf_CSI}
I(X;Y|H)=\log_2{S}-\mathbb{E}_{X,Y,H}\left\{\log_2\frac{\sum_{x\in{A_X}}p(Y|H,x)}{p(Y|H,X)}\right\}.
\end{equation}
The expectation terms in \eqref{eq:Mut_inf} and \eqref{eq:perf_CSI} can be calculated by Monte Carlo simulations.

 \end{document}